\documentclass[traditabstract]{aa}
\usepackage{graphicx}
\usepackage{txfonts}
\usepackage[FIGBOTCAP]{subfigure}
\usepackage{booktabs}
\newcommand{\Rin}{R_{\rm in}}

\newcommand{\Tmax}{T_{\rm max}}
\newcommand{\Tin}{T_{\rm in}}

\newcommand{\rchsq}{\chi_{\nu}^2}
\newcommand{\Rlag}{R_{\rm \tau_K}}

\newcommand{\Rhalf}{R_{\rm 1/2}}
\newcommand{\RVh}{R_{\rm V0.5}}
\newcommand{\swavehalf}{\Lambda_{0.5}}

\newcommand{\tauf}{f_{\varepsilon}}

\newcommand{\Edit}{}
\newcommand{\LE}{}

\newcommand{\miniwidth}{0.3\textwidth}
\newcommand{\figwidth}{6.5cm}
\newcommand{\figintvl}{0.03\textwidth}

\usepackage{natbib}
\begin{document}
%
\title{Mapping the radial structure of AGN tori}


\author{Makoto Kishimoto$^1$,
       Sebastian F. H\"onig$^2$,
       Robert Antonucci$^2$, 
       Florentin Millour$^3$, 
       Konrad R.~W.~Tristram$^1$,\\
       \and
       Gerd Weigelt$^1$
       }

\institute{
           $^1$Max-Planck-Institut f\"ur Radioastronomie, Auf dem H\"ugel 69,
           53121 Bonn, Germany;
           \email{mk@mpifr-bonn.mpg.de}\\
           $^2$Physics Department, University of California, 
  Santa Barbara, CA 93106, USA \\
           $^3$Observatoire de la C\^ote d Azur, 
           Departement FIZEAU, Boulevard de l'Observatoire, BP 4229, 06304, Nice Cedex 4, France
          }


\authorrunning{Kishimoto et al.}

\titlerunning{Radial structure of AGN tori}

 
\abstract{ 

  We present mid-IR interferometric observations of six type 1 AGNs at
  multiple baseline lengths ranging from 27 m to 130 m, reaching high
  angular resolutions up to $\lambda/B \sim 0.02$ arcseconds.  For two
  of the targets, we have simultaneous near-IR interferometric
  measurements as well, taken within a week.  We find that all the
  objects are partially resolved {\LE at} long baselines in these IR
  wavelengths.  The multiple-baseline data directly probe the radial
  distribution of the material on sub-pc scales.  We show that for our
  sample, which is small but spans over $\sim$2.5 orders of magnitudes
  in the UV/optical luminosity $L$ of the central engine, the radial
  distribution clearly and systematically changes with luminosity.
  
  \quad The brightness distribution at a given mid-IR wavelength seems
  to be rather well described by a power law, which makes a simple
  Gaussian or ring size estimation quite inadequate.  In this case, a
  half-light radius $\Rhalf$ can be used as a representative size.  We
  show that the higher luminosity objects become more compact in
  normalized half-light radii $\Rhalf/\Rin$ in the mid-IR, where
  $\Rin$ is the dust sublimation radius empirically given by the
  $L^{1/2}$ fit of the near-IR reverberation radii.  This means that,
  contrary to previous studies, the physical mid-IR emission size
  (e.g. in pc) is not proportional to $L^{1/2}$, but increases with
  $L$ much more slowly. With our current datasets, we find that
  $\Rhalf \propto L^{0.21\pm0.05}$ at 8.5~$\mu$m, and $\Rhalf$ nearly
  constant at 13~$\mu$m.

  \quad The derived size information also seems to correlate with the
  properties of the total flux spectrum, in particular the smaller
  $\Rhalf /\Rin$ objects having bluer mid-IR spectral shape.  We use a
  power-law temperature/density gradient model as a reference, and
  infer that the radial surface density distribution of the heated
  dust grains at a radius $r$ changes from a steep $\sim$$r^{-1}$
  structure in high luminosity objects to a shallower $\sim$$r^{0}$
  structure in those of lower luminosity.  The inward dust temperature
  distribution does not seem to smoothly reach the sublimation
  temperature -- on the innermost scale of $\sim$$\Rin$, a relatively
  low temperature core seems to co-exist with a slightly distinct
  brightness concentration emitting roughly at the sublimation
  temperature.

}

\keywords{Galaxies: active, Galaxies: Seyfert, Infrared: galaxies, 
Techniques: interferometric}

\maketitle
%

\section{Introduction}\label{sec_intro}

A key step in understanding the accretion process in active galactic
nuclei (AGN) is to determine the role and structure of the putative
obscuring torus.  {\LE The obscuration here essentially refers to} the
radiative absorption by dust grains in the torus, which hides the
UV/optical emission of the central engine and the broad line emission
(see \citealt{Antonucci93} for a review).  In type 2 objects, which do
not display these spectral components, our line of sight towards the
nucleus is thought to be relatively close to the system's equatorial
direction along which the torus obscuration becomes effective. Type 1
objects are believed to be rather face-on and permit us to directly
observe the central components unobscured.  The torus dust grains
thermally re-emit the absorbed energy, and the innermost dust
sublimates at $\sim$1500~K, setting the inner boundary of the dust
distribution. This dust sublimation region thus radiates mostly in the
near-infrared (near-IR; wavelength $\lambda$ $\sim$ a few $\mu$m),
with outer dust grains emitting at lower temperatures in the mid-IR
($\sim$10~$\mu$m). Without much spatial information, the torus
structure has long been inferred from the observed spectral energy
distributions (SEDs) mainly in the near- and mid-IR, as well as
theoretical arguments.

The SEDs were first modeled by the tori with smooth density
distributions \citep{Pier93,Granato94,Efstathiou95}. However, the case
for the material being in a clumpy medium has theoretically been
argued from an early stage \citep{Krolik88}.  Accordingly, there have
been many efforts to develop clumpy torus models in the past decade
\citep{Nenkova02,Dullemond05,Hoenig06,Schartmann08,Nenkova08model,Hoenig10model},
and the torus structure has been {\LE inferred} in the context of these
clumpy torus models
\citep{Nenkova08obs,RamosAlmeida09,Mor09,Hoenig10obs,RamosAlmeida11,Alonso-Herrero11}.
However, we are finally beginning to {\LE have} far more {\it direct}
information with the long-baseline interferometry in {\it both} the
near- and mid-IR by spatially resolving the inner structure.

\subsection{Early Type 2 investigations}

Detailed interferometric studies in the mid-IR have targeted a few
bright type 2 AGNs, {\LE with the first ones being} NGC1068 and
Circinus \citep{Jaffe04,Tristram07}, using the mid-infrared
interferometric instrument (MIDI) at the Very Large Telescope
Interferometer (VLTI).  These have achieved the first determination of
the overall size and projected orientation of the mid-IR emission
region with two-temperature elliptical Gaussian models.  The AGN
NGC1068 was also successfully observed with the near-IR (2.2 $\mu$m)
long-baseline interferometric instrument VINCI at VLTI
\citep{Wittkowski04}, and followed up in the mid-IR with more detailed
MIDI observations \citep{Raban09}.  One hard aspect is that, since the
nucleus is heavily obscured in type 2 objects, one needs to {\LE
  disentangle} the effects of radiative transfer {\LE in order to}
study the intrinsic structure and distribution of the material.
Another type 2 object, Cen A, was the first radio-loud object to be
studied with the IR long-baseline interferometry
\citep{Meisenheimer07}, which was followed by more detailed
observations by \cite{Burtscher10}.  Interestingly, the corresponding
mid-IR source is found to be dominated by an unresolved source,
{\LE interpreted} to be a synchrotron core, which might, however, also 
complicate {\LE the study of the structure}.

\subsection{Type 1 study in the near-IR}\label{sec_t1nIR}

Studies of type 1 objects are largely unaffected by the
inclination/obscuration.  We can directly scrutinize the inner
intrinsic structure in both the near-IR and mid-IR. In the near-IR,
\cite{Swain03} made the first measurement of the K-band (2.2~$\mu$m)
visibility of the brightest type 1 AGN NGC4151 with the Keck
interferometer (KI), where they {\LE obtained} quite high visibility
(squared visibility $V^2 \sim 0.8$). This result was recently
confirmed by \cite{Kishimoto09KI} and also \cite{Pott10}.
Furthermore, \cite{Kishimoto09KI,Kishimoto11} found similarly high
visibilities for in total eight type 1 AGNs.  They argued that
this is an indication of the marginal and partial resolution of the
dust sublimation region in these type 1 AGNs. Firstly, they marginally
detected a visibility decrease over the baseline length {\LE for} NGC4151, which
{\LE would be} robust evidence {\LE for} a partially resolved structure.
Secondly, its implied size calculated as a geometric thin-ring radius
for NGC4151 as well as those for the other type 1s are found to
roughly match independent radius measurements from near-IR 
reverberations, which are thought to be probing the dust sublimation
radius.

The near-IR reverberation radius $\Rlag$ is the light-crossing
distance {\LE over} the time lag between the variabilities in the
UV/optical and near-IR (2.2 $\mu$m). The former emission is thought to
originate from the central accretion disk (AD) and the latter from the
hot dust grains in the sublimation region heated by the AD
(e.g. \citealt{Glass04}; \citealt{Minezaki04}; \citealt{Suganuma06};
\citealt{Koshida09}).  The radius $\Rlag$ has been shown to be
proportional to $L^{1/2}$ where $L$ is the AGN luminosity
(\citealt{Suganuma06}; our {\LE specific} definition of $L$ is given
in Sect.\ref{sec_res}), thus, the interferometric ring radii at K-band
{\LE described above} also roughly scale with $L^{1/2}$.  This
is what we would expect for the dust sublimation radii
(e.g. \citealt{Barvainis87}).  In addtion, the measured ring radii
have been shown to be almost unaffected by the near-IR AD component
\citep{Kishimoto07,Kishimoto09KI}.  Therefore, the case for the
near-IR visibilities probing the dust sublimation region is strongly made.

Furthermore, \cite{Kishimoto09KI,Kishimoto11} pointed out that the
ring radii derived from the observed visibilities are {\it either}
roughly equal to {\it or} slightly larger than the $L^{1/2}$-fit to
the reverberation radii as a function of $L$.  Since reverberation
measurements generally place weight on the small {\LE responding}
radii (e.g. \citealt{Koratkar91}), $\Rlag$ is expected to represent the
radius very close to the inner boundary of the dust distribution. On
the other hand, the interferometric ring radii {\LE rather} correspond to
brightness-weighted effective radii.  In this sense, the $L^{1/2}$-fit
reverberation radius gives a good normalization for the radial extent,
taking out the simple $L^{1/2}$ scaling. This normalization is
used extensively in this paper.

\subsection{Type 1 study in the mid-IR}

In the mid-IR, \cite{Beckert08} presented the first long-baseline
mid-IR interferometry for a type 1 AGN, NGC3783.  \cite{Kishimoto09}
published two-baseline data for the same galaxy, revising the results
shown in \cite{Beckert08} with dedicated software (see more details
below) and also adding one more data set taken at a different
baseline.  They proposed a generic way of implementing a uniform
comparison of the interferometric data from different targets with
different luminosities and at different distances, by using the inner
radius $\Rin$ described above for normalizing the spatial scale probed
by the inteferometer.  Mainly using the data of NGC3783, but
supplementing them with those from other type 1s and the type 2 AGN
NGC1068, they argued that the radial brightness distribution in the
mid-IR of type 1 AGNs seems to follow a relatively shallow
power-law. This was a first attempt to directly map the radial
distribution in type 1 AGN tori.

Subsequently, \cite{Burtscher09} also presented two-baseline MIDI data
for another type 1 AGN, NGC4151, and measured the mid-IR emitting
region size using a model of a Gaussian plus point-source, which gives
a visibility curve relatively similar to that of a power-law at low spatial
frequencies.  \cite{Tristram09} presented the results of a MIDI
snap-shot survey for AGNs of both types 1 and 2, and discussed the
mid-IR emission size as a function of luminosity, which has recently
be further evaluated with additional data by \cite{Tristram11}.

Here we present a first systematic investigation of type 1 AGNs with
the mid-IR interferometer, expanding the study of \cite{Kishimoto09}.
Combining our mid-IR data with the near-IR interferometry and
photometry, we aim to comprehensively map the radial structure of the
thermally emitting dusty region in the mid- and near-IR. We aim to
derive physical constraints as directly as possible from the
observations, with the ultimate goal of finding whether the radial
structure is systematically dependent on the properties of the central
engine.

\begin{table*}
\caption[]{Properties of our targets and their point-source flux from our UKIRT images or 2MASS images.} 
{\tiny
\begin{tabular}{llcccccccccccccccccccc}
\hline
\hline
name    & $z^a$    & scale$^b$ & $E_{B-V}^c$ & date & \multicolumn{5}{c}{flux (mJy)$^d$} & $A_V$ & \multicolumn{2}{c}{$\Rlag$ fit$^e$ ($\equiv$ $\Rin$)} & $\nu L_{\nu}(5500\AA)^f$ \\
\cmidrule(rl){6-10} \cmidrule(rl){12-13} 
        & corr. & (pc/mas) & (mag)     & (UT) & Z & Y & J & H & K                       & (mag)        & (pc) & (mas) & (erg s$^{-1}$)$^b$ \\ 
\hline
NGC4151         & 0.00414 & 0.0855& 0.028 & 2009-06-17 & 39.7 & 42.4 & 59.4 & 96.3 & 173. &             & 0.033 & 0.383 & $(8.3\pm1.3)\cdot 10^{42}$\\ 
NGC3783$^g$     & 0.0108  & 0.221 & 0.119 & 1999-04-11 &  --- &  --- & 19.1 & 33.5 & 62.5 &             & 0.061 & 0.274 & $(2.8\pm0.4)\cdot 10^{43}$\\
ESO323-G77$^g$  & 0.0159  & 0.324 & 0.101 & 1999-05-03 &  --- &  --- & 28.6 & 63.4 & 107. & 1.2$\pm$0.4 & 0.10  & 0.321 & $(8.4\pm1.9)\cdot 10^{43}$\\
H0557-385       & 0.0342  & 0.681 & 0.043 & 2009-12-13 & 4.03 & 4.43 & 7.21 & 15.3 & 36.8 & 1.5$\pm$0.5 & 0.12  & 0.181 & $(1.2\pm0.3)\cdot 10^{44}$\\ 
IRAS09149$^{g,h}$ & 0.0579  & 1.12  & 0.182 & 2000-01-05 &  --- &  --- & 35.3 & 59.1 & 116. &             & 0.47  & 0.419 &$(1.7\pm0.2)\cdot 10^{45}$\\
IRAS13349$^i$    & 0.108   & 1.98  & 0.012 & 2009-06-17 & 7.75 & 8.71 & 12.6 & 26.3 & 62.3 &0.93$\pm$0.31 & 0.49  & 0.248 &$(1.9\pm0.4)\cdot 10^{45}$\\ 
\hline
\end{tabular}
\\
$^a$ CMB corrected value from NED. 
$^b$ $H_0=70$ km s$^{-1}$ Mpc$^{-1}$, $\Omega_{\rm m}=0.3$, and $\Omega_{\Lambda}=0.7$.
$^c$ Galactic reddening from \cite{Schlegel98}. 
$^d$ Flux measurement uncertainty is $\sim$5 \% (see Sect.3 in \citealt{Kishimoto09KI}).
$^e$~{\Edit The K-band reverberation radius} $\Rlag$ from UV luminosity using the fit by \cite{Suganuma06}, {\Edit which is defined as $\Rin$ in this paper (see Eq.\ref{eq_rin}).}
$^f$~Optical luminosity at rest wavelength 5500\AA.
$^g$~From 2MASS images. $^h$~IRAS09149-6206. $^i$~IRAS13349+2438.

}
\label{tab_sed}
\end{table*}

\begin{table}

\caption[]{MIDI observation log}
{\tiny
\begin{tabular}{lllccccccccccccc}
\hline
\hline
\multicolumn{2}{c}{date \& time} & telescope & $B_{p}$ & PA        & vis./flux & max\\
\multicolumn{2}{c}{(UT)}         &           & (m)     & ($\degr$) & calibrator & S/N$^a$ \\
\hline
\multicolumn{3}{l}{\bf NGC4151} \\ %
2009-05-09 & 01:20 & UT1-UT3 &  61.5 &  55.9 & HD94264 & 4.1\\
2009-05-09 & 01:32 & UT1-UT3 &  64.1 &  55.7 & HD94264 & 3.3\\
2009-05-09 & 02:37 & UT1-UT3 &  77.3 &  53.0 & HD98262 & 3.3\\
2009-05-09 & 02:49 & UT1-UT3 &  79.3 &  52.3 & HD98262 & 3.4\\
2009-05-10 & 00:44 & UT1-UT2 &  26.7 &  45.1 & HD94264 & 5.1\\
2009-05-10 & 00:55 & UT1-UT2 &  28.1 &  46.1 & HD94264 & 4.9\\
2009-05-11 & 00:33 & UT1-UT4 & 102.0 &  84.8 & HD113996 & 4.1\\
2009-05-11 & 00:44 & UT1-UT4 & 105.1 &  83.1 & HD113996 & 4.3\\
2009-05-11 & 01:19 & UT1-UT4 & 113.4 &  78.3 & HD94264 & 4.2\\
2009-05-11 & 01:31 & UT1-UT4 & 115.8 &  76.7 & HD94264 & 4.0\\
2009-05-12 & 01:59 & UT1-UT4 & 121.3 &  72.4 & HD94264 & 3.6\\
2009-05-12 & 02:28 & UT1-UT4 & 125.2 &  68.3 & HD98262 & 3.3\\
\hline
\multicolumn{3}{l}{\bf NGC3783} \\
2005-05-28 & 03:39 & UT1-UT2 &  44.4 &  45.1 & HD100407 & 3.1\\
2005-05-28 & 04:00 & UT1-UT2 &  42.6 &  45.9 & HD100407 & 5.6\\
2005-05-31 & 03:01 & UT2-UT4 &  68.7 & 114.8 & HD100407 & 3.2\\
2005-05-31 & 03:21 & UT2-UT4 &  64.9 & 119.6 & HD100407 & 3.7\\
2009-05-11 & 03:31 & UT1-UT4 & 108.3 &  80.4 & HD112213 & 3.7\\
2009-05-11 & 23:19 & UT1-UT4 & 128.4 &  45.0 & HD101666 & 4.5\\
2009-05-11 & 23:33 & UT1-UT4 & 129.0 &  47.3 & HD101666 & 4.2\\
\hline
\multicolumn{3}{l}{\bf ESO323-G77} \\
2009-05-08 & 01:01 & UT3-UT4 &  56.8 &  94.4 & HD112213 & 2.1$^b$\\
2009-05-08 & 01:06 & UT3-UT4 &  57.3 &  95.3 & HD112213 & 2.5$^b$\\
2009-05-10 & 03:06 & UT1-UT2 &  54.1 &  30.5 & HD112213 & 3.8\\
2009-05-10 & 03:18 & UT1-UT2 &  53.8 &  31.8 & HD112213 & 3.4\\
2009-05-10 & 04:09 & UT1-UT2 &  51.8 &  37.1 & HD112213 & 3.8\\
2009-05-10 & 04:21 & UT1-UT2 &  51.3 &  38.1 & HD112213 & 4.1\\
2009-05-11 & 04:07 & UT1-UT4 & 118.4 &  75.0 & HD112213 & 4.2\\
2009-05-12 & 00:05 & UT1-UT4 & 126.9 &  36.9 & HD112213 & 3.9\\
2009-05-12 & 00:16 & UT1-UT4 & 127.6 &  39.1 & HD112213 & 3.6\\
\hline
\multicolumn{3}{l}{\bf H0557-385}\\
2009-05-09 & 23:33 & UT1-UT2 &  42.1 &  46.4 & HD40808 & 3.6\\
2009-05-09 & 23:45 & UT1-UT2 &  41.0 &  46.8 & HD40808 & 4.0\\
2009-08-02 & 09:14 & UT3-UT4 &  32.9 &  60.3 & HD9362 & 3.8\\
2009-08-03 & 09:02 & UT1-UT2 &  56.3 & -12.9 & HD39425 & 3.7\\
2009-08-05 & 09:48 & UT1-UT4 & 120.0 &  17.4 & HD40808 & 3.5\\
\hline
\multicolumn{3}{l}{\bf IRAS09149-6206} \\
2009-03-15 & 00:29 & UT1-UT2 &  49.9 &  13.0 & HD80007 & 3.9\\
2009-03-15 & 00:39 & UT1-UT2 &  49.7 &  14.7 & HD80007 & 3.8\\
2009-05-08 & 00:25 & UT3-UT4 &  62.4 & 126.4 & HD89682 & 2.3$^b$\\
2009-05-08 & 01:45 & UT3-UT4 &  62.4 & 144.2 & HD89682 & 2.5$^b$\\
2009-05-08 & 02:01 & UT3-UT4 &  62.3 & 148.0 & HD89682 & 2.0$^b$\\
2009-05-09 & 00:18 & UT1-UT3 &  80.2 &  53.4 & HD89682 & 3.6\\
2009-05-09 & 00:30 & UT1-UT3 &  78.9 &  55.4 & HD89682 & 4.2\\
2009-05-10 & 23:38 & UT1-UT4 & 119.2 &  77.4 & HD89682 & 4.2\\
2009-05-10 & 23:49 & UT1-UT4 & 118.0 &  79.8 & HD89682 & 4.6\\
\hline
\multicolumn{3}{l}{\bf IRAS13349+2438} \\
2009-05-10 & 01:51 & UT1-UT2 &  34.7 &  28.6 & HD113996 & 3.2\\
2009-05-10 & 02:30 & UT1-UT2 &  38.2 &  33.6 & HD113996 & 3.2\\
2009-05-10 & 04:40 & UT1-UT2 &  49.3 &  39.3 & HD112213 & 4.6\\
2009-05-11 & 02:40 & UT1-UT4 & 116.2 &  71.1 & HD113996 & 4.5\\
2009-05-11 & 02:51 & UT1-UT4 & 118.7 &  70.3 & HD113996 & 4.9\\
2009-05-12 & 01:06 & UT1-UT4 &  89.9 &  76.2 & HD113996 & 3.5\\
2009-05-12 & 01:18 & UT1-UT4 &  93.9 &  75.6 & HD113996 & 3.3\\
2009-05-12 & 04:11 & UT1-UT4 & 129.3 &  63.7 & HD127093 & 5.1\\
\hline
\end{tabular}
\\
$^a$ Maximum signal-to-noise ratio (SNR) per smoothed frame for the
correlated\\ flux integraged over the whole N-band (see Appendix 
for details).\\
$^b$ Excluded from the analyses in this paper due to low SNR.
}

\label{tab_obs}
\end{table}

\section{Observations and data reductions}\label{sec_obs}

\subsection{MIDI}\label{sec_midi}

We observed six type 1 AGNs listed in Table~\ref{tab_sed} on 8-12 May
2009 (UT) using MIDI, which is a two-beam combiner working in the
mid-IR wavelengths (\citealt{Leinert03}).  The observation log for
each fringe track is listed in Table~\ref{tab_obs}.  We also collected
a limited number of data sets for two of the six targets on 15 Mar and
5 Aug 2009 (Table~\ref{tab_obs}). The adaptive optics system MACAO at
each of the 8.2m Unit Telescopes was locked on the nucleus of each
target AGN.  A single baseline was allocated for each night, and we
used four different baselines with their projected length spanning
from 27 m to 129 m. For one of the targets, NGC3783, we also used in
our analysis MIDI data taken in 2005 published by \cite{Beckert08} and
\cite{Kishimoto09}.

The data were reduced using the software EWS (version 1.7.1;
\citealt{Jaffe04SPIE}), but with modifications using our own IDL
codes. {\Edit The old NGC3783 data were also re-reduced with the same
  procedure.}  The details of the process are described in the
Appendix~\ref{sec_software}. Briefly, we firstly implemented an
additional background subtraction for the photometry frames using
adjacent sky strips in the 2D spectrum.  Secondly, to reduce errors in
group delay determinations, we smoothed delay tracks over typically
$\sim$20-40 frames. To avoid {\LE possible} positive biases in the
correlated flux, we also averaged $\sim$20-40 frames to determine the
phase offsets.  The system visibility was obtained from the
observations of visibility calibrators taken right after or before
each target observation with a similar airmass.  These data were also
reduced with similar smoothings above to calibrate out the effect of
the time averaging.  These calibrators are also the photometric
standards found in the list by \cite{Cohen99IR} and/or by R. van
Boekel (priv. communication).  We obtained the correlated flux, which
was corrected for the system visibility, and total flux separately.
The errors for the total flux and correlated flux were firstly
estimated from the fluctuation of the measurements over time.  The
total flux spectrum for each object was then derived as the weighted
mean of many measurements over different nights (for NGC3783, the
spectrum was derived separately for 2005 and 2009 data; see below).
For the correlated flux measurements at adjacent $uv$ points, we also
took weighted means. Finally, for the correlated flux, we assigned a
systematic uncertainty as a 5\% of the total flux based on the
analysis presented in Appendix~\ref{sec_software}.

For NGC3783, the mid-IR total flux apparently increased by $\sim$20\%
from 2005 to 2009 observations. To facilitate a direct comparison, we
scaled the total flux and correlated flux from the 2005 observations
by the same factor. This means that we assumed no visibility change in
the mid-IR over this period. This is conceivable based on a comparison
of near-IR interferometry data for NGC4151 that found no significant
visibility change over a one-year timescale (\citealt{Pott10};
\citealt{Kishimoto11}).

To test the accuracy of our MIDI reduction software specifically for
targets at sub-Jy flux levels, we observed stars with flux
$\sim$0.3--0.8 Jy at 12 $\mu$m during our May 2009 run, and also in
Mar 2009. Details are described in Appendix~\ref{sec_software}.  One
significant issue is that, if the frames are not averaged over a
sufficiently long time-interval, the correlated flux is affected by a
positive bias {\LE apparently} in the wavelength region for which
there are lower (intrinsic) counts.  Since MIDI count spectra often
have a peak at shorter wavelengths, the resulting biased spectrum is
often redder than it should be.  In addition, coherence loss toward
shorter wavelengths might not be fully compensated for by calibrator
frames, leading also to redder correlated-flux spectra.  With the data
for the sub-Jy unresolved stars, we demonstrate that the dedicated
codes can correctly estimate the correlated flux, and quantify the
possible systematic uncertainty.

In Table~\ref{tab_obs}, the quality of each fringe track is indicated
by the maximum signal-to-noise ratio (S/N) per smoothed frame (see
Sect.\ref{sec_maxsnr}). The analysis of the sub-Jy unresolved stars
described above currently covers this S/N down to $\sim$2.5, thus in
this paper we decided not to use data with a S/N lower than
2.5. Accordingly, five fringe tracks in Table~\ref{tab_obs} were
excluded from our analysis.

\subsection{The near-IR data}

We combined the mid-IR MIDI data above with data in the near-IR to
comprehensively map the radial structure of the thermally emitting
dusty region.  For the photometry in the near-IR, we have
contemporaneous imaging data at five broad-band wavelengths from 0.9
to 2.2 $\mu$m for NGC4151, IRAS13349, and H0557-385, taken with the
Wide Field Camera (WFCAM) on the United Kingdom Infrared Telecope
(UKIRT).  The data for the former two objects were published in
\cite{Kishimoto09KI}, but, together with the new data for H0557-385,
we present here the data for all three AGNs in Table~\ref{tab_sed}.
The wide field-of-view of WFCAM provides simultaneous
point-spread-function (PSF) measurements {\LE from} stars in the same
field. We used them to implement accurate PSF + host galaxy fits, and
measured the point-source-only SEDs in the near-IR (see
\citealt{Kishimoto09KI} for more details).  For the three other
targets in the sample, we measured the point-source flux by applying
the same procedure to 2MASS images.

We have the near-IR interferometry data taken with the Keck
interferometer (KI) at 2.2 $\mu$m for two of the six MIDI targets,
NGC4151 and IRAS13349. The KI data were taken on 15 May 2009
\citep{Kishimoto09KI}, less than a week from our MIDI observations of
these targets. {\Edit For the general comparison over the sample
  between the mid- and near-IR,} we also use the sample of in total
eight type 1 AGNs, taken with the KI in the near-IR at 2.2 $\mu$m
\citep{Kishimoto09KI,Kishimoto11}.

\subsection{VISIR}\label{sec_visir}

To obtain accurate photometry and check the accuracy of the total flux
spectra from MIDI, we took mid-IR images of each object in two
passbands, namely at 8.6 $\mu$m and 11.9 $\mu$m with VISIR on UT3.
These images were mostly taken within about four weeks of the MIDI
observations, to minimize the effect of variability.  The observation
log and the results are shown in Table~\ref{tab_visir}.  For
ESO323-G77, a VISIR spectrum was taken by \cite{Hoenig10obs} within
several days of our MIDI observations.

All the VISIR images were taken in 'perpendicular' mode with each
dataset consisting of four image tiles.  Using the data from the ESO
pipeline (version 3.7.2), we implemented additional subtraction of the
background determined at annuli of radii $\sim1.5-3.0''$ from each
source on each of the four tiles. After checking the stability of the
flux calibration as a function of aperture radii, we determined the
flux conversion factor at a radius $\sim$$1''$ where its uncertainty
was estimated from the scatter in the results from the four tiles.

\subsection{Additional photometric data in the mid-IR and near-IR}

When fitting the models described in the next section to the SEDs, we
only used our MIDI total flux spectra and WFCAM (or 2MASS)
point-source fluxes described above.  However, we gathered in addition
IRS spectra and IRAC imaging data from the Spitzer Heritage Archive,
listed in Table~\ref{tab_spitzer}, for all the targets.  For NGC3783,
we also have VLT/ISAAC data taken on 2 Jan 2011 (H\"onig et al. in
prep; Table~\ref{tab_spitzer}).  These extra data points from Spitzer
and ISAAC fall approximately on our model SEDs presented below, with
possibly some additional flux from the host galaxy, thus are viewed as
approximate upper limits, and not used in the spectral fitting. We
summarize the usage of our data in Table~\ref{tab_summary}.

\begin{table}

\caption[]{The results of our VISIR imaging observations.}
\begin{tabular}{lllccccccccccccc}
\hline
\hline
name & flux  & aperture$^a$ & UT date & calibrator \\ 
     & (mJy) & (arcsec)     &         &            \\
\hline
\multicolumn{5}{l}{\bf PAH1 at $\lambda$ 8.59 $\mu$m, $\Delta\lambda$ 0.42 $\mu$m} \\ %
NGC4151    & 860$\pm$17 & 1.8 & 2009-06-07 & HD120933\\
NGC3783    & 423$\pm$24 & 1.5 & 2009-06-07 & HD145897\\
H0557-385  & 330$\pm$18 & 1.5 & 2009-09-07 & HD41047\\
IRAS09149  & 346$\pm$10 & 1.5 & 2009-06-07 & HD99167\\
IRAS13349  & 389$\pm$10 & 1.5 & 2009-06-07 & HD145897\\
\multicolumn{5}{l}{\bf PAH2ref2 at $\lambda$ 11.88 $\mu$m, $\Delta\lambda$ 0.37 $\mu$m}\\%
NGC4151    & 1432$\pm$53 & 2.7 & 2009-06-07 & HD120933\\
NGC3783    &  760$\pm$17 & 1.5 & 2009-06-07 & HD145897\\
H0557-385  &  404$\pm$29 & 1.5 & 2009-09-07 & HD41047\\
IRAS09149  &  474$\pm$31 & 1.5 & 2009-06-07 & HD99167\\
IRAS13349  &  530$\pm$26 & 1.5 & 2009-06-08 & HD145897\\
\hline
\end{tabular}
\\
$^a$ Aperture diameter for target and calibrator 
for flux determination.

\label{tab_visir}
\end{table}

\begin{table*}

\caption[]{Spizer data from Spitzer Heritage Archive and ISAAC/VLT data.}
\begin{tabular}{lcccccccccccccccccccccccc}
\hline
\hline
name & IRS & \multicolumn{5}{c}{IRAC flux (mJy)}   \\
\cmidrule(rl){3-7}
& date (UT) & date (UT) & 3.6 $\mu$m & 4.5 $\mu$m & 5.7 $\mu$m & 7.9 $\mu$m  \\ 
\hline
NGC4151   & 2004-01-08 & 2004-12-17 & 264.9$\pm$1.0 & 363.8$\pm$1.2 & 554.0$\pm$3.2 & 924.7$\pm$2.3 \\
NGC3783   & 2007-06-27 & 2007-08-07 & 118.0$\pm$0.7 & 158.5$\pm$0.8 & 228.1$\pm$2.3 & 359.6$\pm$1.5 \\
ESO323-G77& 2006-08-05 & 2009-03-14 & 185.6$\pm$0.8 & 215.4$\pm$1.0 & 288.3$\pm$2.4 & 411.2$\pm$1.6 \\
H0557-385 & 2007-10-06 & 2008-10-31 & 140.7$\pm$0.7 & 195.3$\pm$0.9 & 268.8$\pm$2.4 & 344.0$\pm$1.5 \\
IRAS09149 & 2008-07-02 & 2008-05-13 & 214.4$\pm$0.9 & 261.9$\pm$1.1 & 330.9$\pm$2.6 & 391.8$\pm$1.6 \\
IRAS13349 & 2005-06-07 & 2009-02-02 & 199.1$\pm$4.2 & 242.6$\pm$1.0 & 300.1$\pm$4.2 & 431.4$\pm$1.7 \\
\hline
name & \multicolumn{4}{c}{ISAAC flux (mJy)} \\
\cmidrule(rl){2-5}
     & date (UT) & 2.25 $\mu$m & 3.80 $\mu$m & 4.66 $\mu$m \\
\hline
NGC3783   & 2011-01-02 & 66.21$\pm$0.61 & 107.00$\pm$0.49 & 152.83$\pm$4.65  \\
\hline
\end{tabular}

\label{tab_spitzer}
\end{table*}

\begin{table*}
\caption[]{Summary of the data used for each target.}
\begin{tabular}{lcccccccccccccccccccccccc}
\hline
\hline
name  & \multicolumn{4}{c}{data used for modeling}            
      & \multicolumn{2}{c}{additional photometric data only for} & polar-\\
\cmidrule(rl){2-5}
      & \multicolumn{2}{c}{near-IR} & \multicolumn{2}{c}{mid-IR} 
      & \multicolumn{2}{c}{confirmation or approximate upper limits} & axis$^a$ \\
\cmidrule(rl){2-3} \cmidrule(rl){4-5} \cmidrule(rl){6-7}
      & 1-2.2 $\mu$m flux & interferometry & total flux & interferometry & 3-15 $\mu$m & mid-IR&PA(\degr)\\
\hline
NGC4151   & WFCAM/UKIRT & Keck & MIDI spectrum & MIDI     & IRAC / IRS & VISIR images         & 91$^b$\\
NGC3783   & 2MASS       & --   & MIDI spectrum & MIDI     & ISAAC / IRAC / IRS & VISIR images & 136$^c$\\
ESO323-G77& 2MASS       & --   & MIDI spectrum & MIDI     & IRAC / IRS & VISIR spectrum       & 174$^d$\\
H0557-385 & WFCAM/UKIRT & --   & MIDI spectrum & MIDI     & IRAC / IRS & VISIR images         & 33$^e$\\
IRAS09149 & 2MASS       & --   & MIDI spectrum & MIDI     & IRAC / IRS & VISIR images         & --\\
IRAS13349 & WFCAM/UKIRT & Keck & MIDI spectrum & MIDI     & IRAC / IRS & VISIR images         & 35$^f$\\
\hline
\end{tabular}
\\
$^a$The PA of the polar axis of the system inferred from optical polarization and radio axis. \\
$^b$\cite{Martel98,Mundell03}. $^c$\cite{Smith02}. $^d$\cite{Schmid03}. $^e$\cite{Brindle90}. $^f$\cite{Wills92}.

\label{tab_summary}
\end{table*}

\subsection{Reddening corrections}\label{sec_deredden}

We corrected the SED for Galactic reddening using values from
\cite{Schlegel98} listed in Table~\ref{tab_sed}. We used the reddening
curves of \cite{Cardelli89}, supplemented with that of \cite{Chiar06}
at long wavelengths. We also corrected the SED for the possible
reddening within the host galaxy for IRAS13349, ESO323-G77, and
H0557-385, using the $A_V$ values listed in Table~\ref{tab_sed}.  For
H0557-385, in terms of the optical continuum slope, Balmer decrement,
and host galaxy inclination (\citealt{Winkler92II,Fairall82}), the
reddening is likely to be between those of ESO323-G77 and IC4329A,
which is another Seyfert 1 galaxy {\LE known to be} strongly affected
by host reddening.  For the latter two objects, the estimates of $A_V$
from the different methods exist including that derived from the
gradients of flux variation \citep{Winkler92I}, and are $\sim$1 and
$\sim$2, respectively \citep{Winkler92I}. Without any additional solid
information about the reddening value for H0557-385, we roughly infer
that $A_V=1.5 \pm 0.5$.


\section{Results}

\subsection{The whole data sets and prerequisites}\label{sec_res}

Figures \ref{fig_nfn_uv_sfreq_low} and \ref{fig_nfn_uv_sfreq_high}
present all the data for each of the six targets.  The total and
correlated flux spectra are shown in the top panel, the visibility as
a function of spatial frequency in the middle, and the $uv$ coverage
in the bottom.  In the top panel, all the additional flux data
described in the previous section are also plotted.  The MIDI data
were binned by seven pixels ($\Delta\lambda \sim 0.4 \mu$m) with the
error in each bin taken as the median of the errors {\LE over} the binned
spectral channels (since errors over adjacent channels are
correlated).

For the visibilities, the spatial frequency is shown in units of
cycles per $\Rin$.  As described in Sect.~\ref{sec_t1nIR}, as the most
empirical measure of the inner radius $\Rin$ of the dust distribution,
we adopt the $L^{1/2}$ fit to the K-band reverberation radius $\Rlag$
given by \cite{Suganuma06}.  Here $L$ is the UV luminosity of the
central engine, which is defined as a scaled optical luminosity, $L=6
\nu L_{\nu}(5500\AA)$ \citep{Kishimoto07}, assuming a generic SED over
the UV/optical wavelengths. Here $\nu$ designates the frequency in the
rest frame, and $L_{\nu}$ is the isotropic luminosity per frequency.
The formula for the $L^{1/2}$ fit by \cite{Suganuma06} is given in
\cite{Kishimoto07} as
\begin{equation}
  \Rin \equiv \Rlag = 0.47 \cdot \left( 
  \frac{6\nu L_{\nu}(5500\AA)}{10^{46} \ {\rm erg/sec}}
  \right)^{1/2} \ {\rm pc}.
  \label{eq_rin}
\end{equation}

If the dust temperature structure is primarily determined by the
illumination from the central source, the radial temperature structure
scales with the sublimation radius $\Rin$.  In this case, by
normalizing a length scale with $\Rin$, we can place various
interferometric data on a uniform scale \citep{Kishimoto09}.  A given
configuration of an interferometer probes a certain spatial frequency,
or a certain spatial scale represented by a spatial wavelength, which
is the reciprocal of the spatial frequency.  Here we normalize this
spatial wavelength with $\Rin$ (see the upper axis of the middle panel
of Figs.\ref{fig_nfn_uv_sfreq_low} and
\ref{fig_nfn_uv_sfreq_high}). This corresponds to writing the spatial
frequency in units of cycles per $\Rin$.  With this normalization, we
can eliminate the simple luminosity scaling {\LE originating from}
$\Rin$ being proportional to $L^{1/2}$, as well as a simple distance
scaling.  Different objects with different luminosities and at various
distances have different angular sizes {\LE for} $\Rin$, but the data
can be uniformly viewed when the probed spatial scale is normalized
with $\Rin$.  We can {\LE further} directly determine whether there is
a luminosity dependence in the structure beyond the simple $L^{1/2}$
scaling.

Following the same idea, the total and correlated fluxes can be
normalized using $\Rin^2$. They are shown in the top panel of
Figs.\ref{fig_nfn_uv_sfreq_low} and \ref{fig_nfn_uv_sfreq_high} as
$(\nu L_{\nu} /4 \pi)/ \Rin^2$ and have the same dimension as the
surface brightness in the rest frame of each target.  This is equal to
$\nu_{\rm obs} f_{\nu_{\rm obs}} \cdot (1+z)^4 / \theta_{\rm in}^2$,
where $\nu_{\rm obs}$, $f_{\nu_{\rm obs}}$, and $\theta_{\rm in}$ are
the frequency, observed flux, and angular size of $\Rin$ in the
observed frame, respectively.  The correlated flux in the top panel is
color-coded with the spatial wavelength, while each visibility data
point in the middle panel is color-coded with the (radiation)
wavelength in the rest frame, as indicated by the color bars.  The
total flux, correlated flux, and visibility spectra are shown in
Appendix~\ref{sec_conv_format} using more conventional units to
facilitate direct comparisons with the data in the literature.

\renewcommand{\miniwidth}{0.3\textwidth}
\renewcommand{\figwidth}{6.5cm}
\renewcommand{\figintvl}{0.03\textwidth}

\newcommand{\froot}{nfn_uv_sfreq3_}
\newcommand{\labelroot}{fig_nfn_uv_sfreq_}

\newcommand{\objnamea}{NGC4151}
\newcommand{\objnameb}{NGC3783}
\newcommand{\objnamec}{ESO323-G77}

\begin{figure*}
\input{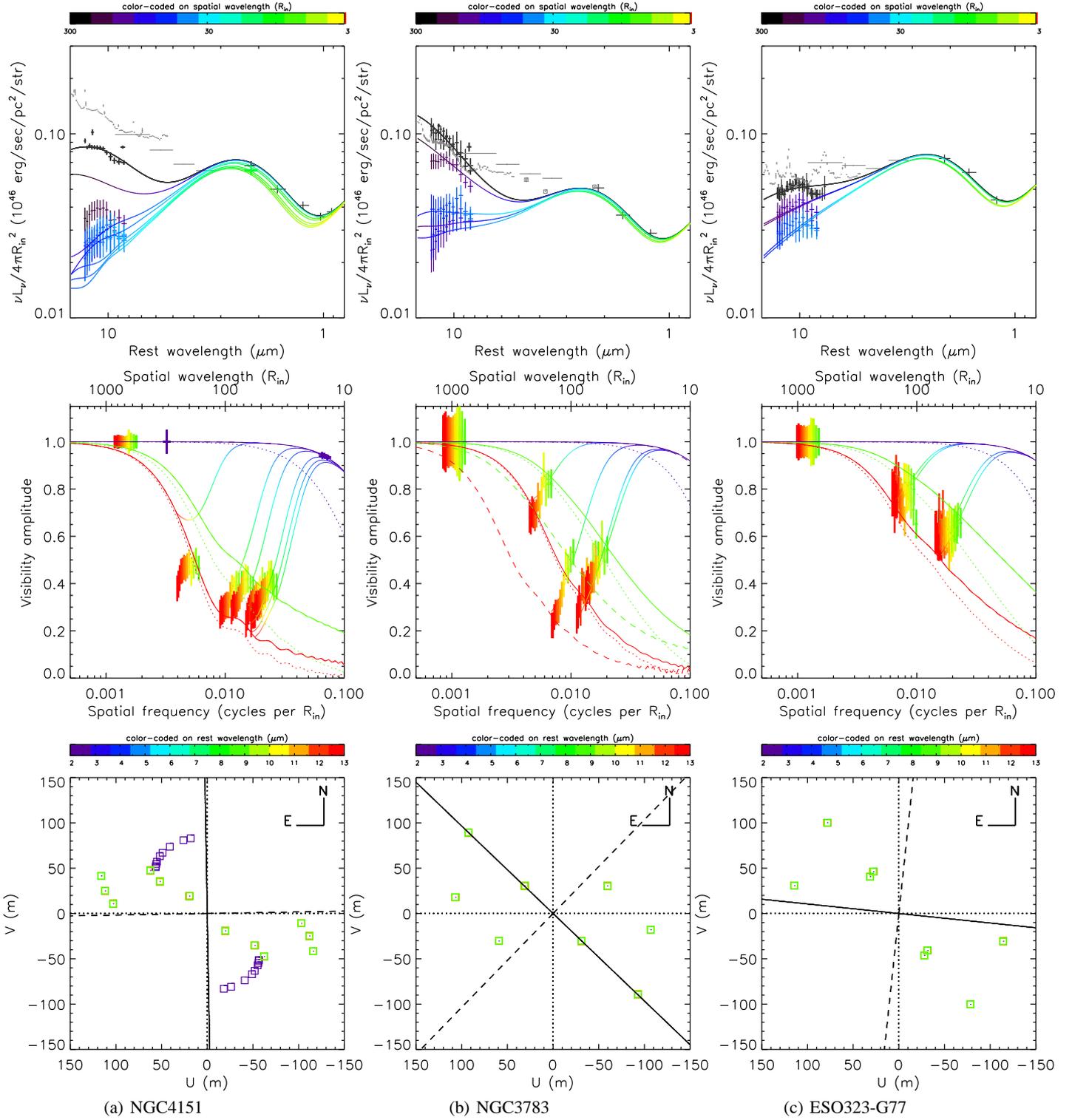}
\caption{The observed data for (a) \objnamea, (b) \objnameb, and (c)
  \objnamec \ (see Sect.\ref{sec_obs}), and the best fits {\LE using} a
  temperature/density gradient model plus an inner 1400K ring (see
  Sect.\ref{sec_powerlaw} and Table~\ref{tab_powerlaw}).  {\it Top}:
  Spectral energy distribution in $(\nu L_{\nu}/4\pi)/\Rin^2$. The
  data plotted in black are the total flux measurements from MIDI and
  VISIR in the mid-IR and from WFCAM or 2MASS in the
  near-IR. Additional total flux spectra and photometry from Spitzer
  (Table~\ref{tab_spitzer}) are plotted in gray. For NGC3783, ISAAC
  photometry data (Table~\ref{tab_spitzer}) are shown as gray squares.
  Correlated fluxes from the mid-IR and near-IR interferometry are
  plotted with colors coded on spatial wavelengths in units of
  $\Rin$ as shown in the top color bar. The model curves for the total
  flux and the correlated fluxes at baselines corresponding to each
  MIDI observation are drawn in the same color coding.  {\it Middle}:
  Visibility amplitude plotted against spatial frequency in units of
  cycles per $\Rin$. The color of the data points corresponds to the
  observing wavelength in the rest frame of the target. This
  color-coding is shown as a color bar between the middle and bottom
  panels.  The iso-wavelength model visibility curves are shown for
  13, 8.5, and 2.2$\mu$m, as well as the curves for each baseline
  configuration (which are 'orthogonal' to the iso-wavelengh curves).
  Dotted curves are the visibilities only for the power-law component,
  with the difference from solid curves indicating the effect of the
  near-IR ring. For NGC3783, solid and dashed model curves are for
  equatorial and polar axis PAs, respectively (see Sect.\ref{sec_PA}).
  {\it Bottom}: The $uv$ coverage of the mid-IR and near-IR
  interferometry with the same color-coding on observing wavelengths.
  Solid and dashed lines indicate the PA of the projected equatorial
  and polar axes, respectively, inferred from optical polarization and
  radio data (see Table~\ref{tab_summary}).}
  \label{\labelroot low}
\end{figure*}

\renewcommand{\objnamea}{H0557-385}
\renewcommand{\objnameb}{IRAS09149}
\renewcommand{\objnamec}{IRAS13349}

\begin{figure*}
\input{fig_nfn_uv_sfreq}
\caption{The same as Fig.\ref{fig_nfn_uv_sfreq_low}, but for (a)
  \objnamea \ (b) \objnameb \ (c) \objnamec.}
\label{\labelroot high}
\end{figure*}

\subsection{MIDI total flux spectra}
\label{sec_totflx}

With MIDI, total fluxes are measured separately and independently of
correlated flux measurements. We derived a total flux spectrum for
each object as a weighted mean of many measurements over different
nights.  The fluxes from our VISIR photometry (or VISIR spectrum in
the case of ESO323-G77) were found to be either in excellent agreement
(within a few percent) with or slightly larger ($\sim$10$-$20\%; see
Appendix~\ref{sec_conv_format}) than the mean MIDI total flux spectrum
for each object. Since the latter is from the beams corrected with
MACAO and {\LE from} a smaller extraction window (adaptive mask in
EWS), it is conceivable that {\LE it has less contribution from} the
extended emission, e.g. from the host galaxy.  In the case of NGC4151,
which seems to have quite an extended mid-IR emission (see the low
visibilities in Fig.\ref{fig_nfn_uv_sfreq_low}a), some of the extended
emission in the VISIR flux is probably excluded in the MIDI flux.  All
these comparisons with VISIR fluxes confirm the accuracy of the
mean MIDI total flux, and its error might be slightly
over-estimated in some cases.

\begin{figure*}
\centering \includegraphics[width=13cm]{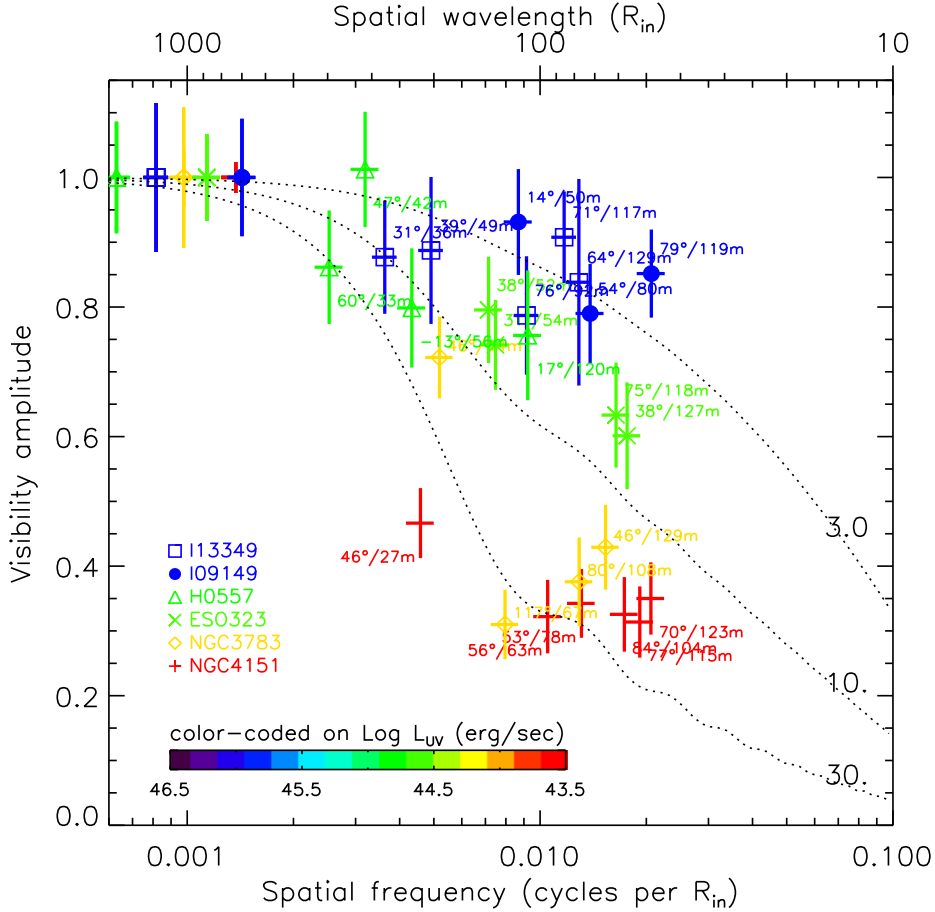}
\caption{Observed mid-IR visibilities of all the targets, each
  averaged over 10 to 12 $\mu$m in the rest frame of each object,
  shown as a function of spatial frequency in units of cycles per
  $\Rin$ (log-scale).  The data are color-coded in terms of the UV luminosity
  of the central engine for the corresponding object, which is defined
  to be 6~$\nu L_{\nu}$(5500\AA) \citep{Kishimoto07}. Each data point
  is plotted with different symbols for different objects, and is
  labeled with the PA / length of the baseline.  It is quite clear
  that the radial structure in units of $\Rin$ changes from being
  compact in higher luminosity objects to extended in lower
  luminosity objects.  Dotted curves are visibility functions for
  power-law brightness distribution with the normalized half-light
  radius $\Rhalf/\Rin$=3, 10, and 30., which roughly depict the
  brightness distributions from higher to lower luminosity objects.}
\label{fig_all_vis_sfreq}
\end{figure*}

In the middle panel for visibilities in
Figs.\ref{fig_nfn_uv_sfreq_low} and \ref{fig_nfn_uv_sfreq_high} (and
also in Fig.\ref{fig_all_vis_sfreq} discussed in the next section), we
show the {\LE visibility uncertainties originating from} each
correlated flux measurement without the contribution of total flux
errors, but show the total flux errors separately as data points of
visibility unity at the lowest spatial frequency.  In this way, we can
clearly show the relative uncertainties between each correlated flux
measurement and the total flux measurement.  In addition, these total
flux data points are shown at the spatial frequency with the baseline
length equal to the single telescope diameter.  As long as the source
remains unresolved by a single telescope, the total flux gives the
correlated flux at this spatial frequency, and the visibility is unity
at this spatial frequency.  Thus, the data points illustrate the
constraint in such a case, which is expected to hold at least
approximately for our sample of type 1 objects. We note however that
these are only for the data presentation.  For the model fitting in
Sect.\ref{sec_rhalf} and \ref{sec_disc}, we include the total flux
error in the visibility error budget in a normal way.

\subsection{Mapping radial structure}

Each panel of Figs.\ref{fig_nfn_uv_sfreq_low} and
\ref{fig_nfn_uv_sfreq_high} shares a common range of axes over
different objects so that we can directly compare them. We can also
place together the data for all the targets on a single plane.
Fig.\ref{fig_all_vis_sfreq} is such a figure, showing the observed
mid-IR visibilities of all the targets, each averaged from 10 to 12
$\mu$m in the rest frame of each object.  In
Appendix~\ref{sec_linear}, we also show the same figure with the
spatial frequency axis on a linear scale.

Our mid-IR observations cover the range of several tens to hundreds
of $\Rin$ in spatial wavelengths. Over this range, the observed
visibilities are relatively concentrated in one locus, from 0.3 to
0.9. If the visibilities of different objects followed the same curve,
this would mean that the size scales only with $L^{1/2}$ and objects
share the same radial structure as inferred by
\cite{Kishimoto09}. However, this is not the case. 

In Fig.~\ref{fig_all_vis_sfreq}, each object is color-coded with its
UV luminosity as defined in Sect.\ref{sec_res}.  It is quite clear
from the figure that within the small sample, which nevertheless spans
over $\sim$2.5 orders of magnitudes in luminosity, the radial
structure does depend on luminosity, i.e. the structure in units of
$\Rin$ looks more compact in higher luminosity objects. This implies
that the increase in the mid-IR physical size (e.g. in pc) with
luminosity is more gradual than $L^{1/2}$. We discuss
this luminosity dependence in Sect.\ref{sec_size_uv}.

While the results above are essentially model-independent, we 
attempt in the following to describe the radial surface-brightness
distribution more specifically with simple models.  The dotted curves
in Fig.~\ref{fig_all_vis_sfreq} are visibility functions for power-law
brightness distributions, which we argue below approximately
depict the observed brightness distributions (see more in
Sect.\ref{sec_rhalf}).

\subsection{Single Gaussian/ring versus power-law}

\begin{figure}
\centering \includegraphics[width=9cm]{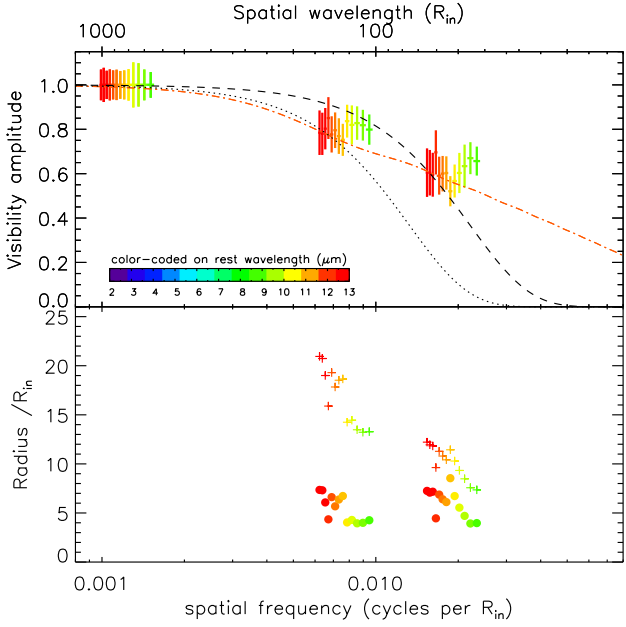}
\caption{{\it Top:} Observed visibilities for ESO323-G77 as a function
  of spatial frequency along the position angle $\sim$$38\degr$.
  Visibility curves for Gaussian distributions with a HWHM/$\Rin$ of
  20 (dotted) and 12 (dashed), and a power-law distribution with
  $\Rhalf/\Rin=7$ (dash-dot) are over-plotted.  {\it Bottom:} Gaussian
  HWHM (plus signs) and $\Rhalf$ (filled circles) are plotted against
  the same spatial frequency. The red to green colors indicate the
  observing wavelengths in the rest frame from 13 to 8.5 $\mu$m. For a
  given wavelength, while the Gaussian HWHM is baseline-dependent, the
  half-light radius is approximately baseline-independent, thus
  seems adequate for representing the whole structure. }
\label{fig_comp_rhalf_ghwhm_obs}
\end{figure}

We first consider the results for ESO323-G77, where we have the data
with the least ambiguity. The total flux spectrum provides with an
excellent match to the simultaneous VISIR spectrum (see
Sect.\ref{sec_visir}; Fig.\ref{fig_fnu_vis_low}) and we have the two
datasets at almost exactly the same PA of the baselines
($\sim$$38\degr$) with good S/N.  The visibility data are shown in
Fig.\ref{fig_comp_rhalf_ghwhm_obs} as a function of spatial frequency.

As in the middle panel of Figs.  \ref{fig_nfn_uv_sfreq_low} and
\ref{fig_nfn_uv_sfreq_high}, the visibilities in
Fig.\ref{fig_comp_rhalf_ghwhm_obs} are presented in different colors
for different wavelengths as indicated by the color bar.  For a given
baseline, shorter wavelengths probe the source at higher
  spatial resolutions, and if the overall source size is the same
when seen at different wavelengths, then the visibility is expected to
become lower, i.e. more resolved, toward shorter
wavelengths. However, the observed visibility spectra are roughly
flat, or increase toward shorter wavelengths, which means that the
overall size of the source is smaller at shorter wavelengths.  This is also
generally seen in other objects, except at some objects' shortest
wavelengths where visibilities might be slightly underestimated owing to
a coherence loss (see Appendix~\ref{sec_atmadj}).

On the other hand, when we consider the visibilities at each wavelength
as a function of baselines (or spatial frequencies), we can easily see
that a simple Gaussian or ring 
({\LE which we treat here almost equally, since they give
almost identical visibility curves at low spatial frequencies well
before the first null; see Sect.\ref{sec_gauss} }) 
does not fit them well, as shown by the dotted
and dashed Gaussian curves in  the top panel of
Fig.\ref{fig_comp_rhalf_ghwhm_obs}. If we derive a Gaussian HWHM or
ring radius for each baseline measurement at a fixed wavelength, we
obtain a smaller size for longer baselines as we show in the
  bottom panel of Fig.\ref{fig_comp_rhalf_ghwhm_obs}. We appear to
observe here a brightness distribution where the
corresponding Gaussian/ring sizes become progressively smaller at
longer baselines.  This probably means that there is a continuous
distribution of brightness spread over a wide range of
continuous spatial scales. This can be understood if the brightness
distribution resembles a power-law distribution at a given
wavelength, with its index generally becoming steeper at
shorter wavelengths.  This is what we advocated in
\cite{Kishimoto09} (see their Fig.1).  This power law is at least
partially expected physically, since the temperature of the directly
illuminated dust grains is expected to follow a power-law with the
radius from the illumination source, simply based on thermal
equilibrium (e.g. \citealt{Barvainis87}).

\begin{figure}
\centering \includegraphics[width=9cm]{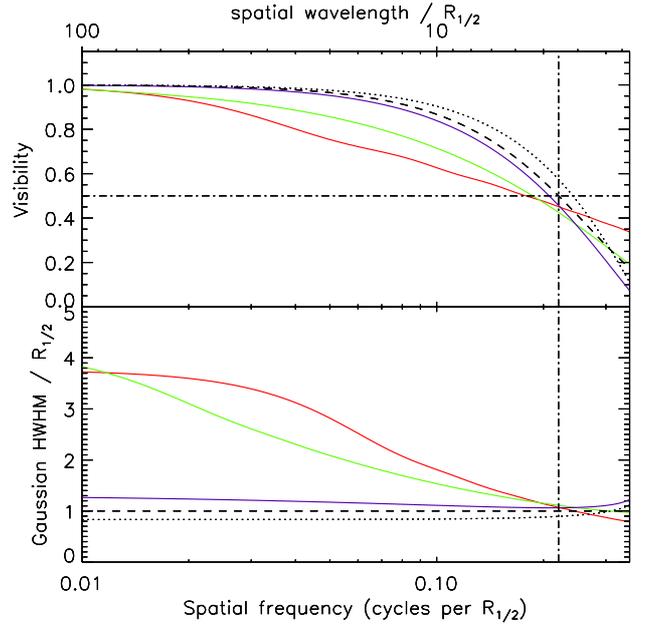}
\caption{Top: Comparison between visibility functions for Gaussian,
  shown as a dashed curve, and power-law distributions where red, green,
  and purple curves correspond to the cases of $\Rhalf$ = 6, 2, 1.3
  $\Rin$, respectively. For comparison, the visibility function of a
  thin ring with radius being equal to the Gaussian HWHM
  (i.e. $\Rhalf$) is shown as a dotted curve.  Bottom: The ratio of
  Gaussian HWHM to $\Rhalf$ is plotted against the same spatial
  frequency in each case. The vertical dot-dash line corresponds to
  the spatial wavelength of 4.5 $\Rhalf$.  }
\label{fig_comp_rhalf_ghwhm}
\end{figure}

\subsection{Half-light radius for power-law brightness distribution}
\label{sec_rhalf}

For given inner and outer radii, a power-law index specifies a
power-law brightness distribution, and a convenient way of
characterizing the corresponding size is to use the half-light radius
$\Rhalf$, the radius within which half of the total integrated light
at a given wavelength is contained.  This can be directly compared
with the HWHM of a Gaussian brightness distribution, since $\Rhalf$ is
equal to the HWHM in a Gaussian case.  When the brightness
distribution is adequately described by a power law, the values of
$\Rhalf$ derived from visibilities at different baselines should all
be the same for a given wavelength. The bottom panel of
Fig.\ref{fig_comp_rhalf_ghwhm_obs} presents the deduced $\Rhalf$ for
each visibility observed. We can see that the deduced $\Rhalf$ is
quite independent of the baseline lengths, in contrast to the Gaussian
HWHM. This reassures that the power-law description in this
case is quite adequate. We can also see that the size becomes smaller
toward shorter wavelengths, as discussed above.

The disadvantage of {\LE adopting a power-law model and its
  corresponding $\Rhalf$} is that we need to define both the inner and
outer radii, which are set here to be 1$\Rin$ and 100$\Rin$,
respectively. However, we can also define another, very similar, more
model-independent radius, the half-visibility radius $\RVh$ as
follows.  A visibility at a given low spatial frequency (within the
first lobe of the visibility function before the first null) roughly
corresponds to the fraction of the flux that is contained within the
spatial scale being probed and remains unresolved by the
interferometer's configuration.  Here the configuration is set by the
observing wavelength $\lambda$ and baseline length $B$.  In this
sense, $\Rhalf$ roughly corresponds to the spatial scale {\LE probed
  by the interferometer for which} the visibility becomes 0.5.  We can
refer to this spatial scale using the spatial wavelength $\Lambda
\equiv \lambda / B$, which is the reciprocal of the spatial frequency for
the configuration, and corresponds to the spatial resolution
of the configuration (\citealt{Kishimoto09}; see Sect.\ref{sec_res}).
It is straightforward to show that the spatial wavelength $\swavehalf$
at which visibility becomes 0.5 for a Gaussian is related to $\Rhalf$
by
\begin{equation}
\Rhalf = \frac{\ln 2}{\pi} \ \swavehalf \ \ {\rm (for \ a \ Gaussian)}.
\end{equation}
If we define the half-visibility radius $\RVh$ as
\begin{equation}
\RVh \equiv \frac{\ln 2}{\pi} \ \swavehalf \simeq \frac{\swavehalf}{4.5},
\end{equation}
then this is equal to $\Rhalf$ for a Gaussian, and can generally be
referred to as a good approximation of the half-light radius
model-independently. This radius can approximately be deduced by
interpolating observations when observed visibilities cross over 0.5
at the longest available baseline.  We can also say more generically
that the visibility at a given spatial wavelength $\Lambda$ within the
first lobe gives the approximate fraction of flux originating from the
region within the radius $\sim$$\Lambda/4.5$.  We note again that for a
Gaussian brightness distribution, all three quantities, namely
HWHM, half-light radius $\Rhalf$ and half-visibility radius $\RVh$,
are identical.

\begin{figure}
\centering \includegraphics[width=9cm]{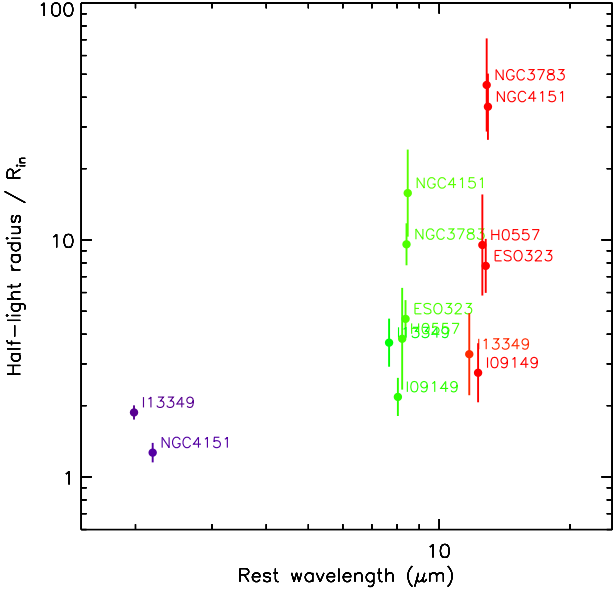}
\caption{Normalized half-light radii $\Rhalf/\Rin$ plotted against
  rest wavelengths. Thin-ring radii at 2.2$\mu$m, normalized by
  $\Rin$, from the KI data for two objects \citep{Kishimoto09KI} are
  also shown.  The normalized radii become smaller for shorter
  wavelengths in each object from the mid-IR to near-IR.}
\label{fig_rhalf_lam}
\end{figure}

\begin{figure}
\centering \includegraphics[width=9cm]{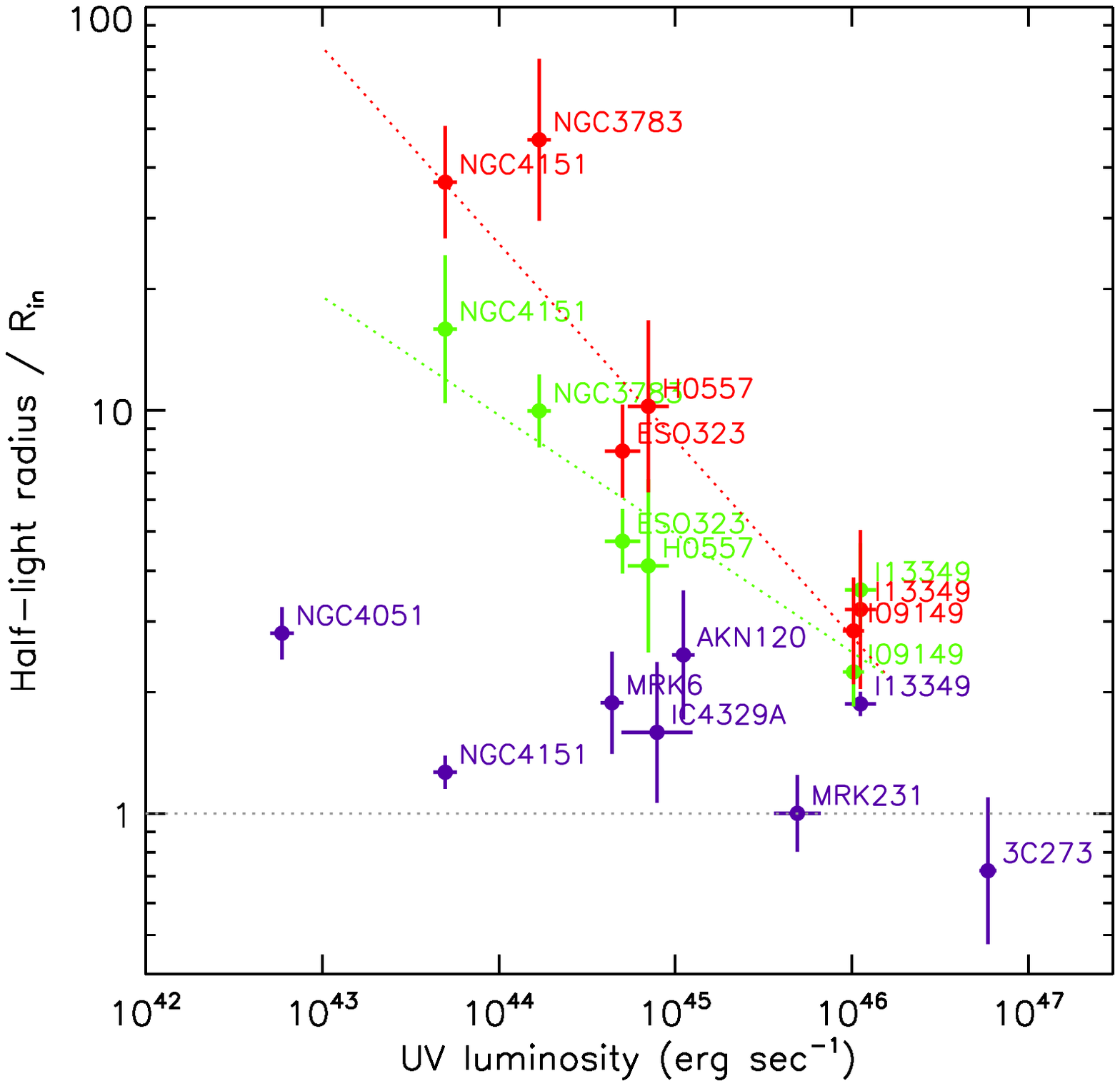}
\caption{Normalized half-light radii $\Rhalf/\Rin$ (red and green for
  13 and 8.5 $\mu$m, respectively), plotted against UV luminosity $L$,
  defined as 6~$\nu L_{\nu}$(5500\AA) \citep{Kishimoto07}.  The red
  and green dotted lines are power-law fits to $\Rhalf/\Rin$ at 13
  $\mu$m and 8.5 $\mu$m, respectively. The normalized radii become
  more compact in the objects having the central engine with a higher
  UV luminosity.  Also shown in purple are thin-ring radii at 2.2
  $\mu$m normalized by $\Rin$ for the sample observed with the KI
  \citep{Kishimoto11}, including two overlapping objects. The gray
  dotted horizontal line indicates radius = $\Rin$. }
\label{fig_rhalf_luv}
\end{figure}

\begin{table*}

\caption[]{Estimated half-light radii $\Rhalf$ in two rest-frame
  wavelengths in the mid-IR and their uncertainties in dex.}
\begin{tabular}{llccccccccccccccccccccccccc}
\hline
\hline
name & $\Rin$$^a$
 & \multicolumn{3}{c}{8.5 $\mu$m} & \multicolumn{3}{c}{13 $\mu$m}\\                                  
\cmidrule(rl){3-5}\cmidrule(rl){6-8} 
          & (pc) & $\Rhalf/\Rin$  & $\Rhalf$ (pc) & unc. (dex) & $\Rhalf/\Rin$  & $\Rhalf$ (pc) & unc. (dex) \\

\hline
NGC4151   & 0.033 &  15.9 & 0.52 & 0.18 & 36.8 & 1.21 & 0.14 \\
NGC3783   & 0.061 &  10.0 & 0.60 & 0.09 & 46.9 & 2.84 & 0.20 \\
ESO323    & 0.10  &  4.7  & 0.49 & 0.08 &  7.9 & 0.83 & 0.12 \\
H0557-385 & 0.12  &  4.1  & 0.51 & 0.21 & 10.2 & 1.26 & 0.21 \\
IRAS09149 & 0.47  &  2.2  & 1.06 & 0.09 &  2.8 & 1.33 & 0.13 \\
IRAS13349 & 0.49  &  3.6  & 1.76 & 0.12 &  3.2 & 1.58 & 0.20 \\
\hline
\end{tabular}
\\
$^a$ See Table~\ref{tab_sed} for $\Rin$ ($\equiv \Rlag$ fit) in mas.

\label{tab_hlight}
\end{table*}

In the top panel of Fig.\ref{fig_comp_rhalf_ghwhm}, we compare the
visibility functions of Gaussian and power-law distributions with
various $\Rhalf$. The curves are shown as a function of spatial
frequency in units of cycles per $\Rhalf$ (or spatial wavelength in
units of $\Rhalf$), so that we can directly compare different $\Rhalf$
power-law cases with the same Gaussian curve. We can see that
power-law curves with smaller values of $\Rhalf$ become
asymptotically more similar to the Gaussian/ring curve.  The bottom
panel shows the ratio of Gaussian HWHM, derived from each visibility,
to $\Rhalf$ as a function of the same spatial frequency.  For extended
power-law distributions, the dependence of the Gaussian HWHM on the
baseline length is quite strong, but for steep distributions, the
dependence on baseline becomes quite small.  All curves cross {\LE the}
visibility $V$ of 0.5 at spatial wavelength $\Lambda$ close to
4.5~$\Rhalf$, meaning that the half-visibility radius $\RVh$ closely
represents the half-light radius $\Rhalf$ in all cases.

The dotted curves in Fig.~\ref{fig_all_vis_sfreq} are the visibility
functions for the power-law brightness distributions with three
different steepnesses, which are quantified by the normalized
half-light radius $\Rhalf/\Rin$=3, 10, and 30 (corresponding to
power-law indices of -2.6, -2.0, and -1.5, respectively). These seem
to closely describe the mid-IR surface brightness distribution over
the sample.  For the more compact structures seen in the higher
luminosity objects, it is more difficult to differentiate the
power-law from a Gaussian/ring for a given measurement
uncertainty. However, based on relatively resolved cases such as in
ESO323-G77, we infer that the power-law description with different
values of $\Rhalf$ over the sample is adequate at least
approximately. We can then translate the observed visibility curves on
the 2D plane (Fig.\ref{fig_all_vis_sfreq}, or the middle panels of
Figs.\ref{fig_nfn_uv_sfreq_low} and \ref{fig_nfn_uv_sfreq_high}) into
a value of $\Rhalf$ at each wavelength, and then disscuss the relation
between this size information and other observables.

We derived the best-fit normalized half-light radius $\Rhalf/\Rin$ for
each object by fitting the visibility curve of a power-law brightness
distribution to the observed visibility data, separately for four
wavelength bins, namely 8.2-10, 10-11, 11-12, and 12-13 $\mu$m in the
observed frame.  To ensure a uniform comparison at the same rest
wavelength over the sample, we fitted a power law as a function of
wavelength over the derived $\Rhalf/\Rin$ (and its error) at the four
wavelength bins, and used this fit to derive values at 8.5 and 13
$\mu$m in the rest frame. The results are summarized in
Table~\ref{tab_hlight}.  The uncertainties are formal 1-$\sigma$
uncertainties for the fits, except for NGC4151 and NGC3783 where the
targets are well-resolved ($V<0.5$ at long baselines) and the
uncertainty in $\Rhalf$ was estimated {\LE as the} dispersion {\LE in}
those derived from different baseline data.  Since the dispersion in
$\Rhalf$ about its best-fit value was more symmetric in log space,
the uncertainties are shown in dex.

We note that the fits were done with the inner and outer radius fixed
at 1 and 100 $\Rin$. The latter can affect the radius estimation for
the extended cases of NGC4151 and NGC3783. However, as long as each
source is at least approximately unresolved by the single telescope,
the half-visibility radius $\RVh$ is quite well-constrained as can be
seen in the middle panels of Fig.\ref{fig_nfn_uv_sfreq_low}$a$ and
Fig.\ref{fig_nfn_uv_sfreq_low}$b$ as well as in
Fig.\ref{fig_all_vis_sfreq} (see Sect.\ref{sec_totflx}). Therefore, our
$\Rhalf$ estimation would also be quite robust for the two targets.

\begin{figure*}
\centering \includegraphics[width=13cm]{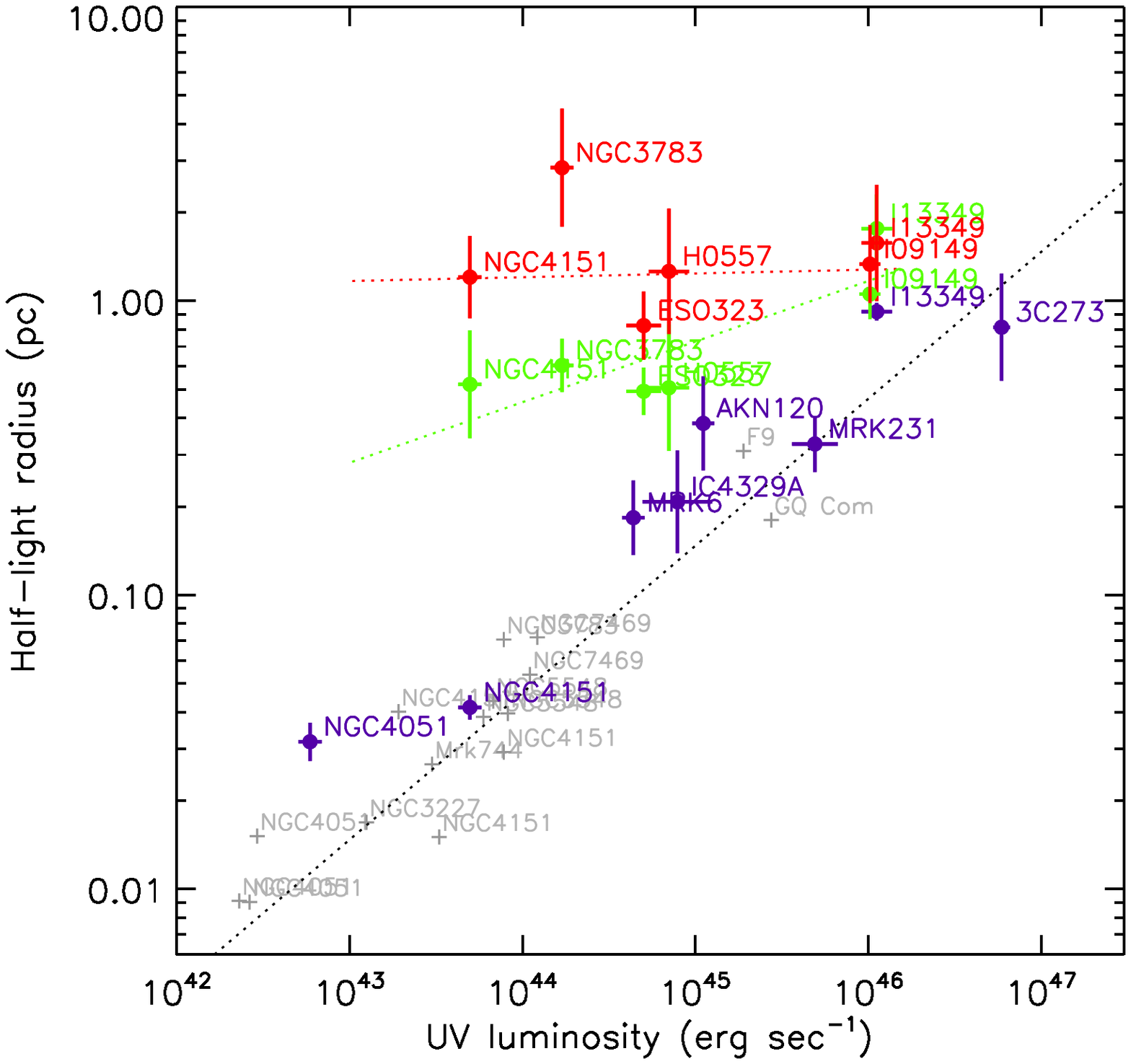}
\caption{Half-light radii $\Rhalf$ in pc (red and green for 13 and 8.5
  $\mu$m, respectively), plotted against UV luminosity $L$, which is
  defined as 6~$\nu L_{\nu}$(5500\AA) \citep{Kishimoto07}.  The red
  and green dotted lines are power-law fit to $\Rhalf$ at 13$\mu$m
  ($\propto L^{0.01\pm0.07}$) and 8.5$\mu$m ($\propto
  L^{0.21\pm0.05}$), respectively. In this physical scale, these
  mid-IR radii increases with luminosity much slower than $L^{1/2}$,
  or almost constant at 13~$\mu$m.  Thin-ring radii at 2.2$\mu$m also
  in pc are plotted in purple for the KI-observed sample. Plotted with
  gray plus signs are the near-IR reverberation radii
  (\citealt{Suganuma06} and references therein), with the black dotted
  line showing their $L^{1/2}$ fit. This fit is the definition of the
  inner radius $\Rin$ in this paper.}
\label{fig_rhalfpc_luv}
\end{figure*}

\section{Discussion}
\label{sec_disc}

\subsection{Sizes at various wavelengths as a function of UV
  luminosity}
\label{sec_size_uv}

Using the half-light radius $\Rhalf$ derived in the previous
section, we discuss below the overall relations of the size to
observing wavelengths and flux-related quantities such as spectral
shape, total flux, and luminosity.

Fig.\ref{fig_rhalf_lam} shows $\Rhalf /\Rin$ as a function of
rest-frame wavelengths.  For the near-IR, we have plotted the
thin-ring radii from the KI data for the two objects in our present
MIDI sample, NGC4151 and IRAS13349 \citep{Kishimoto09KI}.  The
thin-ring approximation should be fairly robust in the near-IR, where
the radial brightness distribution is expected to be much steeper and
the power-law visibility curve becomes very close to that of a
Gaussian/ring (Fig.\ref{fig_comp_rhalf_ghwhm}). A more detailed
estimation of $\Rhalf$ would require knowledge of the shape of the
innermost dust distribution, which is not yet well constrained. Thus,
we simply adopt here the thin-ring approximation for the near-IR data.
We would infer that the other four targets, for which no KI data are
available, probably have near-IR ring radii of $\sim$1$-$2 $\Rin$
based on the KI-observed sample of type 1 AGNs (see
Fig.\ref{fig_rhalf_luv}; \citealt{Kishimoto11}).  From the mid-IR to
the near-IR, the normalized emission size would {\LE then become}
certainly smaller with wavelength, very roughly following some
power law for each object, but with different indices for different
objects.

As we demonstrated in Fig.\ref{fig_all_vis_sfreq}, the mid-IR emission
size seems to have a luminosity dependence.  This is now
quantitatively shown in Fig.\ref{fig_rhalf_luv}, where $\Rhalf/\Rin$
is plotted against UV luminosity $L$.  The data points are color-coded
for the observing wavelengths, and we include the 2.2 $\mu$m thin-ring
radii of all the KI-observed sample from \cite{Kishimoto11}.  As we
can see, the ratio of half-light radius to $\Rin$ becomes smaller with
increasing UV luminosity.  This implies that, for a given radial
temperature distribution, the higher luminosity objects have a steeper
radial density structure, at least in the mid-IR emitting radii, as we
discuss in greater detail below.  We also plot power-law fits to the
$\Rhalf/\Rin$ at 8.5 and 13 $\mu$m in Fig.\ref{fig_rhalf_luv}.

We can also analyze the luminosity dependence of {\it un}normalized
$\Rhalf$ {\LE in} a physical scale.  Fig.\ref{fig_rhalfpc_luv} shows the
half-light radius $\Rhalf$ in pc plotted against the same UV
luminosity.  In the physical scale, the radius in the mid-IR increases
with luminosity much slower than $L^{1/2}$.  Power-law fits to
$\Rhalf$ in pc at {\Edit 8.5 and 13 $\mu$m} as a function of luminosities are
\begin{equation}
  \Rhalf \ (8.5 \mu {\rm m}) = (1.2\pm0.2) \cdot \left( 
  \frac{6\nu L_{\nu}(5500\AA)}{10^{46} \ {\rm erg/sec}}
  \right)^{0.21 \pm 0.05} \ {\rm pc},
  \label{eq_rhalfpc8}
\end{equation}
and
\begin{equation}
  \Rhalf \ (13 \mu {\rm m}) = (1.3\pm0.3) \cdot \left( 
  \frac{6\nu L_{\nu}(5500\AA)}{10^{46} \ {\rm erg/sec}}
  \right)^{0.01 \pm 0.07} \ {\rm pc},
  \label{eq_rhalfpc13}
\end{equation}
{\Edit where we see that $\Rhalf$ at 13 $\mu$m is consistent with being constant.}

These results are {\LE contrary to} those of previous studies of the
mid-IR size as a function of luminosity by \cite{Tristram09} and
\cite{Tristram11}, who concluded that the mid-IR size is consistent
with {\LE being proportional to} $L^{1/2}$.  They used a single
Gaussian for the size estimation, and this can lead to a large
systematic error owing to the baseline dependency.  When the radial
brightness distribution has a relatively shallow power-law form, a
Gaussian size would represent the emission size that is resolved by a
given interferometer configuration, leading to the strong dependence
on the sampled $uv$ points.  Thus, in this case, the simple Gaussian
size is quite inadequate for representing the emission size for the
whole distribution.  Here we account for the effect of the $uv$
sampling more properly, based on the inference that the brightness
distribution is of roughly a power-law form.  This enables us to
perform a more accurate, uniform comparison over the sample, as shown
in Fig.\ref{fig_rhalfpc_luv}.  To {\LE gain} a greater accuracy and
model-independency, we can always return to the uniform view of the
visibility curves over the sample shown in
Fig.\ref{fig_all_vis_sfreq}.

\begin{figure}
\centering \includegraphics[width=9cm]{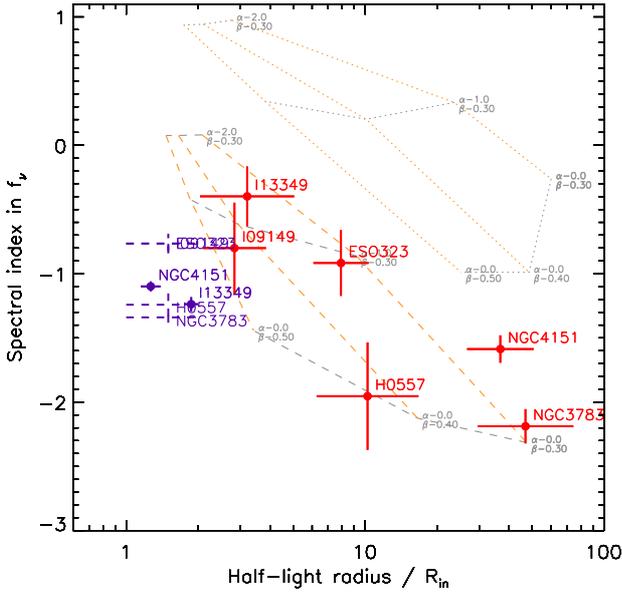}
\caption{The mid-IR (8.5$-$13 $\mu$m) spectral indices in $f_{\nu}$
  are plotted in red against normalized half-light radii $\Rhalf/\Rin$
  at 13 $\mu$m.  Also plotted in purple are spectral indices between
  2.2 and 13 $\mu$m against thin-ring radii at 2.2 $\mu$m normalized
  by $\Rin$. For the four objects without near-IR interferometry data, we
  indicated the near- to mid-IR spectral indices in dashed lines over
  the thin-ring radius range of 1$-$2 $\Rin$, roughly expected based
  on other targets observed with the Keck interferometer (see
  Fig.\ref{fig_rhalf_luv}). The dotted line shows the grid for a
  simple power-law model with radial surface density index $\alpha$ of
  (-2.0, -1.0, -0.0) and temperature index $\beta$ of (-0.5, -0.4,
  -0.3) with the innermost tempeture $\Tin$=1400K. The grid covers the
  observed range of $\Rhalf/\Rin$, but systematically bluer in
  spectral shape. The dashed lines show the same grid but with
  $\Tin$=700K, which apparently matches the mid-IR observations over
  the sample when the density index ranges from $\sim$$-1$ to 0 in
  different objects with a temperature index of $\sim$$-0.35\pm0.05$
  (for details, see Sect.\ref{sec_powerlaw}). }
\label{fig_clr_rhalf}
\end{figure}

The half-light radius in units of $\Rin$ has a quite close
relationship with the mid-IR spectral
shape. Figure~\ref{fig_clr_rhalf} shows $\Rhalf/\Rin$ plotted against
spectral index $a$ in $f_{\nu} \propto \nu^a$ in the mid-IR. The shape
becomes bluer when the size is smaller.  In terms of the
radial structure, this is what we would expect, as the steeper radial
structure looks bluer as the contribution from hotter dust becomes
more dominant. We elaborate on this point later using simple models.

\begin{figure}
\centering \includegraphics[width=9cm]{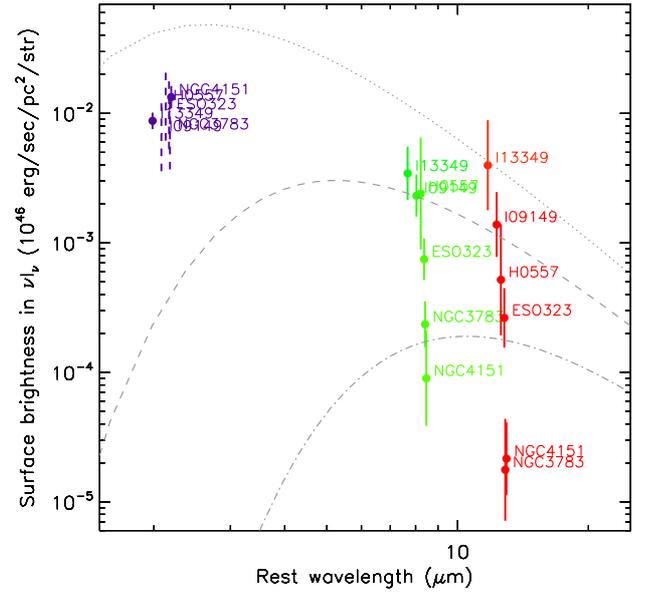}
\caption{Average surface brightnesses $\nu I_{\nu}$ in the rest frame,
  defined as $(\nu L_{\nu}/4\pi) /(\pi \Rhalf^2)$, are plotted against
  rest wavelengths. Overplotted are the Planck functions with
  temperatures 1400, 700, and 350 K (dotted, dashed, and dash-dot,
  respectively). For NGC4151 and IRAS13349, which have near-IR Keck
  interferometry data, the average surface brightness at 2.2 $\mu$m is
  calculated in the same way but using thin-ring radii. For the
  objects without near-IR interferometry, it is calculated assuming
  $\Rhalf$=(1.5$\pm$0.5) $\Rin$ and indicated by dashed lines. See
  further details in Sect.\ref{sec_size_uv}.}
\label{fig_nuinu_lam}
\end{figure}

\begin{figure}
\centering \includegraphics[width=9cm]{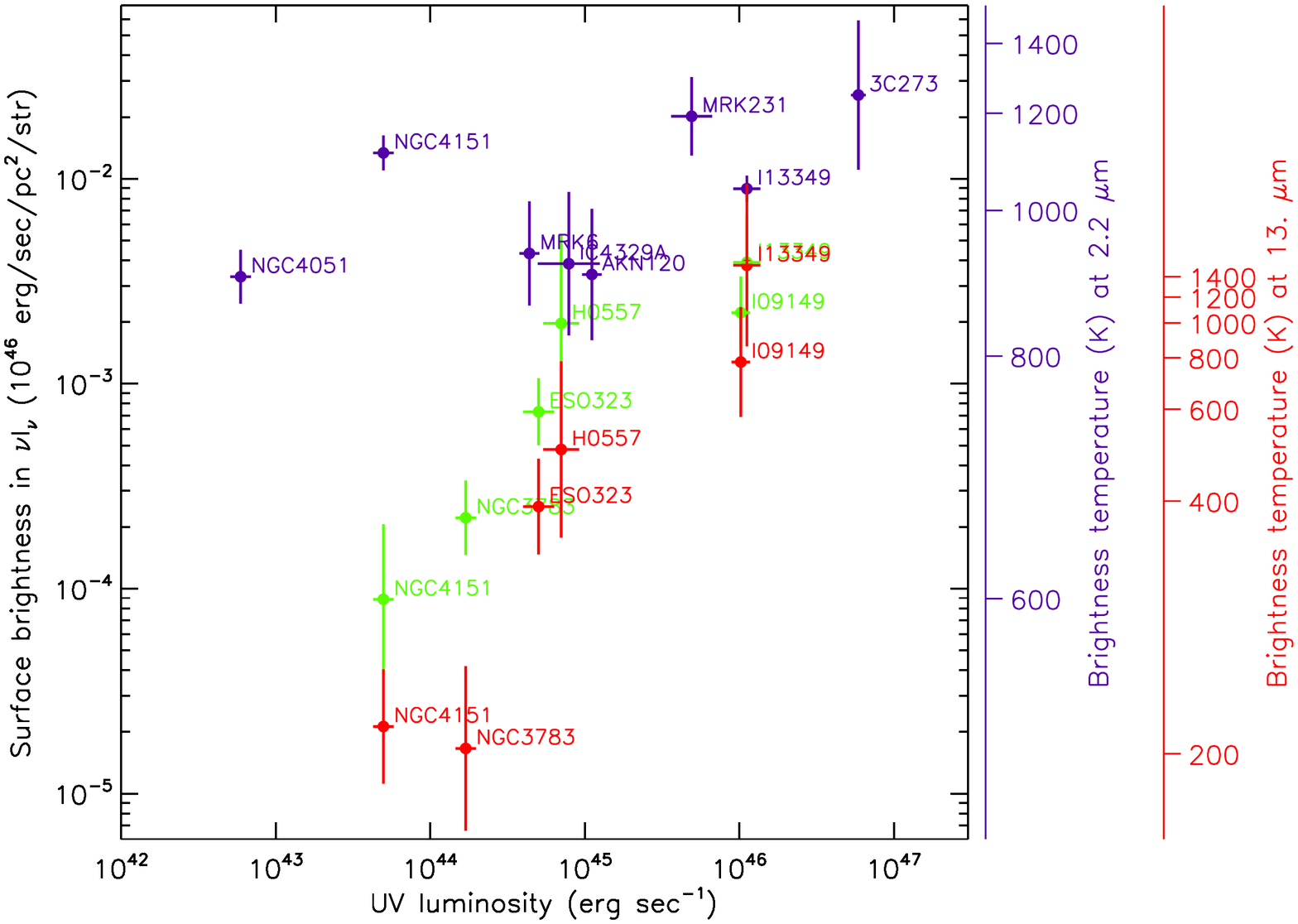}
\caption{The same average surface brightness as in
  Fig.\ref{fig_nuinu_lam}, but plotted against the UV luminosity of
  the central engine, defined as 6~$\nu L_{\nu}$(5500\AA)
  \citep{Kishimoto07}. The red and green symbols correspond to 13 and
  8.5 $\mu$m, respectively. The surface brightness at 2.2 $\mu$m is
  shown in purple for those with the near-IR interferometry data from
  \cite{Kishimoto11}. The luminosity dependence is quite clear at
  least for the surface brightness in the mid-IR.  }
\label{fig_nuinu_luv}
\end{figure}

We can also calculate an average surface brightness in the rest frame
using $\Rhalf$ as $(\nu L_{\nu}/4\pi)/(\pi \Rhalf^2)$, which is
equivalent to $\nu_{\rm obs} f_{\nu_{\rm obs}} \cdot (1+z)^4 /(\pi
\theta_{1/2}^2)$ where $\theta_{1/2}$ is the angular size of $\Rhalf$.
Fig.\ref{fig_nuinu_lam} shows this as a function of rest-frame
wavelengths. This can be directly compared with the Planck function as
shown in the same figure, where the comparison gives the brightness
temperature\footnotemark[1] at a given wavelength. In the near-IR, all
the objects show the brightness temperature $\sim$1400$-$1000K, while
in the mid-IR the brightness temperature ranges from $\sim$1000K to
$\sim$200K. In the near-IR, the brightness temperature is quite close
to the observed color temperature of $\sim$1400K, meaning that the
average surface filling factor at the innermost dusty region must be
close to unity.  Figure~\ref{fig_nuinu_luv} shows the surface
brightness as a function of UV luminosity, and the corresponding
brightness temperatures at 2.2 and 13 $\mu$m are shown in the second
and third y-axis, respectively. We can again clearly see the luminosity
dependence.

The close relationship between $\Rhalf/\Rin$ and the mid-IR spectral
index (Fig.\ref{fig_clr_rhalf}) and the relationship between
$\Rhalf/\Rin$ and UV luminosity (Fig.\ref{fig_rhalf_luv}) implies that
there should also be a correlation between the mid-IR spectral index
and luminosity.  {\LE In fact}, the luminosity dependence of the radial dust
distribution was already inferred by \cite{Hoenig10obs}, based on the
IR spectral comparison between Seyfert galaxies and high-luminosity
type 2 QSOs.  Here we directly show the distribution change by
spatially resolving it.

\footnotetext[1]{Here the brightness temperature is defined without
  requiring the intensity to be in the Rayleigh-Jeans limit.}

\begin{table*}

\caption[]{The results of two-temperature Gaussian fits.}
\begin{tabular}{lllcccccccccccccccccccccccc}
\hline
\hline
name & $\Rin$$^b$
 & \multicolumn{3}{c}{near-IR Gaussian} & \multicolumn{3}{c}{mid-IR Gaussian}                                  & $\rchsq$ $^{(c)}$ & minor/major\\
\cmidrule(rl){3-5}\cmidrule(rl){6-8} 
          & (pc) & T(K)         & HWHM ($\Rin$) & $f_0$         & T(K)        & HWHM ($\Rin$) &$f_0$          &          & \\
\hline
NGC4151   & 0.033 & 1389$\pm$68  & 1.06$\pm$0.02  & 0.36$\pm$0.10 & 282$\pm$6   & 51.1$\pm$1.2 & 0.12$\pm$0.01 &  12.9 \\
NGC3783   & 0.061 & 1220$\pm$108 & (fixed to 1)   & 0.57$\pm$0.28 & 242$\pm$9   & 32.4$\pm$0.6 & 0.77$\pm$0.14 &   8.1     \\
NGC3783$^a$& 0.061 & 1220$\pm$108 & (fixed to 1)   & 0.57$\pm$0.28 & 242$\pm$9   & 39.5$\pm$1.0 & 1.12$\pm$0.21 &   4.1     & 0.46$\pm$0.02 \\
ESO323    & 0.10 & 1539$\pm$175 & (fixed to 1)   & 0.29$\pm$0.14 & 320$\pm$18  & 12.3$\pm$0.4 & 0.71$\pm$0.15 &   1.7 \\
H0557-385 & 0.12 & 1241$\pm$54  & (fixed to 1)   & 0.97$\pm$0.26 & 307$\pm$15  & 18.2$\pm$1.2 & 1.06$\pm$0.26 &   1.4 \\
IRAS09149 & 0.47 & 1204$\pm$103 & (fixed to 1)   & 0.63$\pm$0.32 & 290$\pm$39  &  7.4$\pm$0.5 & 1.67$\pm$0.89 &   0.65 \\
IRAS13349 & 0.49 & 1314$\pm$57  & 1.57$\pm$0.06  & 0.37$\pm$0.10 & 366$\pm$26  &  9.6$\pm$0.7 &  1.9$\pm$0.6  &   0.98 \\
\hline
\end{tabular}
\\
$^a$ An elliptical Gaussian is fit for the mid-IR component, assuming
the major axis at 136\degr. The quoted mid-IR HWHM is along this major
axis.\\ $^b$ See Table~\ref{tab_sed} for $\Rin$ ($\equiv \Rlag$ fit) in mas.
$^c$ Reduced $\chi^2$ values for visibility fits.

\label{tab_gauss}
\end{table*}

\subsection{Multiple-temperature Gaussian fit}\label{sec_gauss}

While the power-law brightness description at each wavelength gives a
relatively model-independent measurement of the emission size, it does
not physically describe the change in the overall size with wavelength
and the flux that we observe.  A simple, though very approximate way
of describing the power-law brightness with changing steepness over
different wavelenths is to consider two or more rings/Gaussians of
different size at different temperatures. At each wavelength, the
multiple components with different sizes contribute differently to the
visibility function, giving a wider range in spatial frequency than
the range covered by a single ring/Gaussian.  The contribution ratio
of the multiple temperature components changes with wavelength,
causing the overall apparent size to change with wavelength.

We {\LE adopted} a Gaussian geometry for each component, though rings
might provide a more physically accurate description of in particular
the innermost dusty region.  In terms of the visibility function, a
ring and Gaussian essentially give the same curve at the low spatial
frequencies before the first null (Fig.\ref{fig_comp_rhalf_ghwhm}; see
e.g. \citealt{Millour08}). Both a thin-ring and a Gaussian geometry
have only one parameter, but a Gaussian permits us to more simply
quantify the average surface brightness of the component, and its
visibility curve (which is also a Gaussian) is the simplest (no nulls,
no lobes).

We describe each Gaussian component as a brightness distribution
$S_{\nu}$ at a radius $r$ from the center given as
\begin{equation}
S_{\nu}(\nu, r) = B_{\nu}(\nu, T) \cdot f_0 \ln 2 \cdot e^{-r^2 \ln 2 /\Rhalf^2}, 
\end{equation}
where $B_{\nu}$ is the Planck function at frequency $\nu$ and
temperature $T$, and $\Rhalf$ is the {\Edit physical} size of the
Gaussian in HWHM, where HWHM = $\Rhalf$ for Gaussian.  The 
  isotropic luminosity $L_{\nu}$ of this component is
\begin{equation}
L_{\nu}/4\pi = f_0 \cdot B_{\nu}(\nu, T) \cdot \pi \Rhalf^2.
\label{eq_gaussflux}
\end{equation}
The factor $f_0$ gives the average surface brightness with respect to
the blackbody when all the emission is contained within HWHM (see more
on emissivity and surface filling factor described in
Sect.\ref{sec_powerlaw}).  In addition to these two Gaussians, we also
include the accretion disk component, which is assumed to remain
unresolved and have a spectrum of $f_{\nu} \propto \nu^{+1/3}$ at IR
wavelengths longward of 0.8 $\mu$m \citep{Kishimoto09KI,Kishimoto08}.

Table~\ref{tab_gauss} shows the results of two-Gaussian-component
fits. We assumed that HWHM=1$\Rin$ for the $\sim$1400K near-IR
component here, except for the two objects for which we have KI data.
We see that the $\sim$300K mid-IR component has a HWHH of several to a
few tens of $\Rin$, which implies an effective temperature radial
gradient of $\beta$ from $\sim$$-0.8$ to $\sim$$-0.4$ where $T \propto
r^{\beta}$. This effective index includes the effect of the radial
gradient of surface density distribution.  In
section~\ref{sec_powerlaw}, we attempt to separate the temperature and
density gradients using more physical but simple models.

\subsection{PA dependence}\label{sec_PA}

The PA coverage of our baselines for each object is very limited.
Since our targets are all type 1 AGNs, we do not expect
to see a large PA dependence in visibilities. Nevertheless, 
we seem to see some PA dependence at least in one object, NGC3783.
This has already been pointed out by \cite{Hoenig10obs}. 
{\Edit The brightness distribution} along the SE-NW direction {\Edit seems}
more extended than that in the NE-SW direction.

We can complement our inner PA dependency study with optical
polarization measurements. The projected direction of the system polar
axis can be inferred to be either parallel or perpendicular to the PA
of the optical continuum polarization, depending on the line
polarization properties \citep{Smith04may,Kishimoto04}. If a linear
radio structure is detected, this direction also {\LE indicates} that
of the system polar axis.  All the targets except IRAS09149 have
optical polarization measurements
\citep{Brindle90,Wills92,Martel98,Schmid03}, and a clear radio axis is
also available for NGC4151 \citep{Mundell03}. The resulting
  inferred PAs are summarized in Table~\ref{tab_summary}, and shown
in the bottom panel of the $uv$ coverage for each object in
Fig.\ref{fig_nfn_uv_sfreq_low} and \ref{fig_nfn_uv_sfreq_high} with
polar and equatorial directions in dashed and solid lines, respectively.

To evaluate the elongation of the mid-IR emission in NGC3783, we
attempted to fit an elliptical Gaussian for the mid-IR component 
with the major axis fixed at the polar axis PA 136\degr, and
the results are listed in Table\ref{tab_gauss}. The emission seems to
be a factor of $\sim$2 elongated along this {\it polar} direction,
suggesting that some of the mid-IR emission originates from the polar
region. More extensive data and analyses will be presented for this
object by H\"onig et al. (2011 in prep).

\begin{figure*}
\centering \includegraphics[width=19cm]{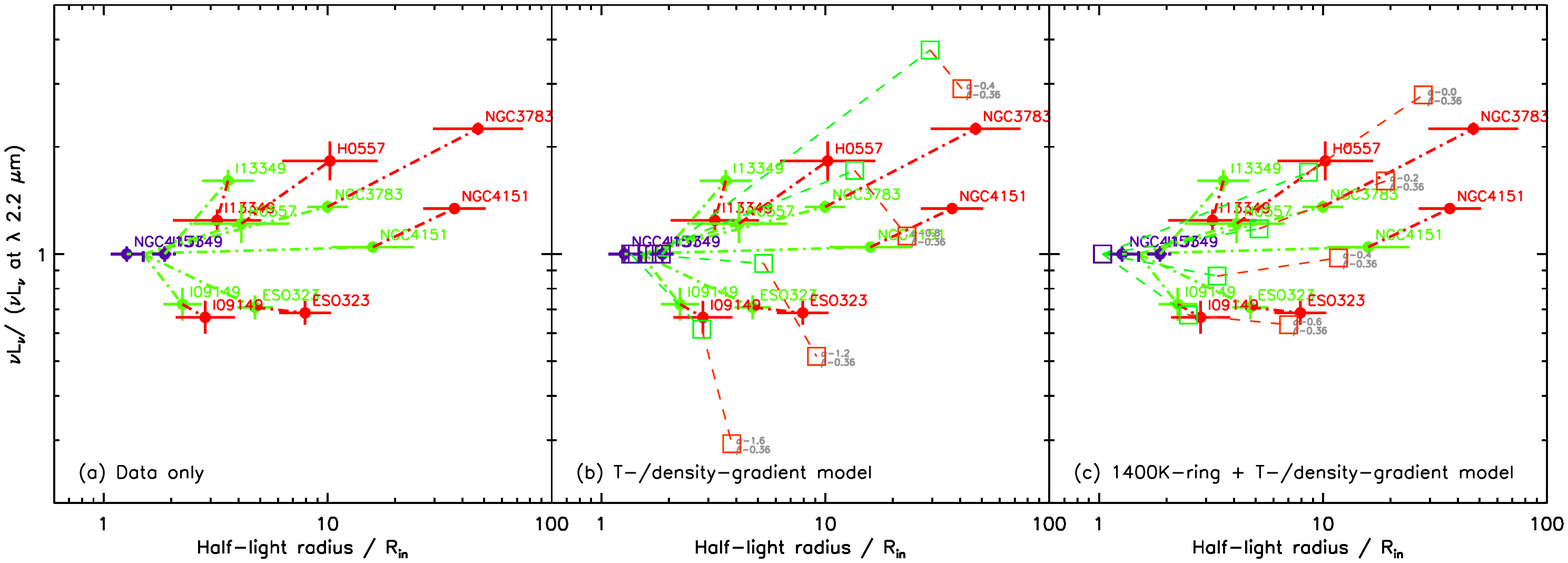}
\caption{(a) SED in $\nu L_{\nu}$ for each object normalized at
  $\lambda$ = 2.2 $\mu$m, plotted against normalized half-light radius
  $\Rhalf/\Rin$, for 2.2, 8.5, and 13 $\mu$m (purple, green, and red,
  respectively).  (b) The same data but with the grid in dashed lines
  for temperature/density gradient models with radial surface density
  indices of $\alpha$=$-$0.4,$-$0.8,$-$1.2,$-$1.6 for the case of $T
  \propto r^{-0.36}$, plotted for the three wavelengths 2.2, 8.5, and
  13 $\mu$m (purple, green, and red squares). This simple power-law
  model does not explain the observed SED and size at the same time.
  (c) The same data but with the grid in dashed lines for the
  power-law (density $\propto r^{-0, -0.2, -0.4, -0.6}$, $T\propto
  r^{-0.36}$) + 1400K thin-ring at $\Rin$, which roughly explain the
  observed SED and size. The two reddened sources IRAS13349 and
  H0557-385 could be interpreted in terms of the same model with a
  slightly larger filling factor for the power-law component.}
\label{fig_normnfn_rhalf_3panel}
\end{figure*}

\begin{table*}

\caption[]{The fit results of temerature/density gradient model + inner ring.}
\begin{tabular}{lcccccccccccccccccccccccc}
\hline
\hline
name & \multicolumn{3}{c}{power-law ($\Tin=700$ K)} & \multicolumn{2}{c}{inner 1400K ring}             & $\chi_{\nu}^2$ & polar / \\
\cmidrule(rl){2-4}\cmidrule(rl){5-6}
          & density index $\alpha$  & $T$ index $\beta$       & $f_0$         & radius ($\Rin$) & $f_{0}$ &              & equatorial\\
\hline
NGC4151   & $-0.31\pm0.04$ & $-0.26\pm0.01$ & $0.16\pm0.01$ & $1.25\pm0.03$ & $0.26\pm0.02$ & 3.5 \\
NGC3783   & $+0.04\pm0.05$ & $-0.31\pm0.01$ & $0.16\pm0.02$ & (fixed to 1)  & $0.29\pm0.02$ & 2.4 \\
NGC3783$^a$&$+0.04\pm0.06$ & $-0.35\pm0.01$ & $0.13\pm0.02$ & (fixed to 1)  & $0.28\pm0.02$ & 0.69 & $2.43\pm0.18$\\
ESO323-G77& $-0.77\pm0.09$ & $-0.30\pm0.02$ & $0.55\pm0.09$ & (fixed to 1)  & $0.42\pm0.03$ & 0.60 \\
H0557-385 & $-0.65\pm0.10$ & $-0.31\pm0.02$ & $1.40\pm0.26$ & (fixed to 1)  & $0.47\pm0.02$ & 1.3 \\
IRAS09149 & $-0.46\pm0.21$ & $-0.42\pm0.03$ & $0.69\pm0.19$ & (fixed to 1)  & $0.27\pm0.02$ & 0.68 \\
IRAS13349 & $-1.26\pm0.16$ & $-0.27\pm0.03$ & $3.6\pm1.0$   & $1.87\pm0.07$ & $0.17\pm0.01$ & 0.71 \\
\hline
\end{tabular}
\\
$^a$ Fit with additional ellipticity for power-law component, assuming polar axis at 136\degr.

\label{tab_powerlaw}
\end{table*}

\subsection{Power-law brightness distribution:
  temperature/density-gradient model}\label{sec_powerlaw}

\subsubsection{Model description}

We now discuss the direct physical constraints on the
radial structure of the inner dusty region thermally emitting in the
IR. To this end, we use a simple {\Edit but generic description of the
  radial surface brightness distribution} with as few parameters as
possible.  To physically describe a face-on radial brightness
distribution $S_{\nu}(\nu, r)$ at radius $r$ and observing frequency
$\nu$, we would need at least two separate distributions, namely
the temperature and surface density distributions.  
{\Edit Here we slightly generalize the surface brightness description in
\cite{Kishimoto09} as}
\begin{equation}
S_{\nu}(\nu, r) = \tauf(r) \cdot B_{\nu}(\nu, \Tmax(r)).
\end{equation}
The idea behind this description is that, at each radius, the face-on
surface brightness is dominated by the emission from the heated dust
grains that have a maximum temperature at that radius
(e.g. \citealt{Hoenig10model}).  These grains are probably at around
the surface of the torus, and most likely directly illuminated by the
central source, if there are such {\LE directly illuminated} grains at
that radius.  Under this maximum-temperature approximation, we refer
to the dust at $\Tmax$ {\LE simply} as heated dust grains/clouds.
We note that our observations are insensitive to much colder, interior
dust grains that may exist deeper along our line of sight at that
radius.\footnotemark[2] We write this maximum temperature $\Tmax(r)$
at radius $r$ as
\begin{equation}
\Tmax=\Tin \cdot (r/\Rin)^{\beta}.
\end{equation}

\footnotetext[2]{There should be very little or no directly
  illuminated dust grains situated at greater depth from the surface
  along our line of sight, because the vertical optical thickness
  (along our line of sight) is {\LE quite likely} smaller than the
  equatorial optical thickness.}

We mainly consider a discrete, clumpy, or inhomogeneous cloud
distribution.  The surface brightness of each heated cloud at radius
$r$ would approximately scale with $r$ in proportion to
$B_{\nu}(\Tmax(r))$.  The additional factor $\tauf(r)$ above, which we
write in the form of a power law as
\begin{equation}
\tauf(r)= f_0 \cdot (r/\Rin)^{\alpha},
\end{equation}
describes the radial run of the surface density (i.e. per unit area)
distribution of these heated discrete clouds (i.e. at $\Tmax$), but
also includes the emissivity of each heated cloud surface. More
specifically, the dimensionless factor $\tauf$ can be regarded as a
surface filling factor multiplied by the emissivity.  In the case of
optically thick illuminated clouds where their emissivity does not
depend sensitively on the radial distance from the illuminating source
or the observing frequency (e.g. \citealt{Hoenig10model}, see their
Fig.3), the factor $\tauf(r)$ is roughly proportional to the radial
surface density distribution of the heated clouds, or heated dust
grains.  Thus, as long as the emissivity is roughly constant over the
radius and frequency, we can regard $\tauf(r)$ as a normalized surface
density of heated dust grains (i.e. again those at $\Tmax$). In this
way, our description is generic enough to ensure that we can obtain as
direct constraints as possible from the observations.

The model would be called a temperature gradient model, if the density
distribution is set to be uniform, i.e. $\alpha=0$, and the
brightness described only by the temperature index $\beta$. In this
case, the index would be an effective temperature index, which would
include the effect of the density gradient.  Here we try to separate
the temperature and density gradient.  At least in the case of the
direct illumination of a dust grain, we would expect that
$\beta$=$-0.5$ for large grains and $\beta$$\sim$$-0.36$ for
standard-size ISM grains (e.g. \citealt{Barvainis87}).  By referring
to these values, we attempt here to obtain constraints on the surface
density distribution.

\subsubsection{Mid-IR spectral slope and SED-size relation}

In Fig.\ref{fig_clr_rhalf}, we show a grid for the two indices
$\alpha$ and $\beta$ on the plane of the mid-IR spectral index and
half-light radius.  The upper-most grid with dotted lines is for
$\Tin=1400$K, which corresponds to the observed color temperature in
the near-IR and is thought to be roughly equal to the dust sublimation
temperature.  The range of density index $\alpha$ from $-$2 to 0 for a
given temperature index reproduces the observed range of normalized
half-light radii $\Rhalf/\Rin$. However, the expected mid-IR spectral
shape as shown by the dotted grid is systematically bluer than
observed.  {\LE In fact}, the observed range of the mid-IR spectral
index, as well as that of $\Rhalf/\Rin$, is reproduced well with the
power-law distribution if $\Tin$ is $\sim700$K, where the
corresponding grid is shown in dashed lines in
Fig.\ref{fig_clr_rhalf}.  We need to accommodate this size-color
constraint.

The total flux spectra, including the mid-IR spectral shape, and the
size information for all the objects can be summarized on the plane of
the SED, normalized at 2.2 $\mu$m where all objects show similar
brightness temperature (Fig.\ref{fig_nuinu_lam}), plotted against
$\Rhalf/\Rin$ as shown in Fig.\ref{fig_normnfn_rhalf_3panel}$a$.  For
objects without Keck near-IR interferometric data, we set their
$\Rhalf$ at 2.2 $\mu$m to have the expected range of values from 1 to
2 $\Rin$ (see Fig.\ref{fig_rhalf_luv}; \citealt{Kishimoto11}).  In the
mid-IR, the objects {\LE show} a relatively uniform behavior. From the
high to low luminosity objects, as the normalized size $\Rhalf/\Rin$
becomes larger, the mid-IR spectral shape becomes gradually redder
(which is already seen in Fig.\ref{fig_clr_rhalf}), and the mid-IR
total flux level gradually increases with respect to the near-IR flux,
although two objects, IRAS13349 and H0557-385, seem to have slightly
(a factor of a few) higher mid-IR flux level than the other four
objects (see more below).

Overplotted in dashed lines in Fig.\ref{fig_normnfn_rhalf_3panel}$b$
on the same data set is the SED-size relation at three wavelengths
expected from the power-law distribution with the temperature index
$\beta$=$-0.36$ over the range of density indices $\alpha$=$-$1.6 to
$-$0.4 (the case with slightly steeper $\beta$ is similar, but with a
slightly different density index range).  The total flux mismatch can
be clearly seen across this plane.  It is quite difficult to reproduce
the wide range of $\Rhalf/\Rin$ and the overall near- to mid-IR SED
which is relatively flat in $\nu f_{\nu}$, while keeping the
relatively red mid-IR spectral shape, for this single power-law
structure.

\begin{figure}
\centering \includegraphics[width=9cm]{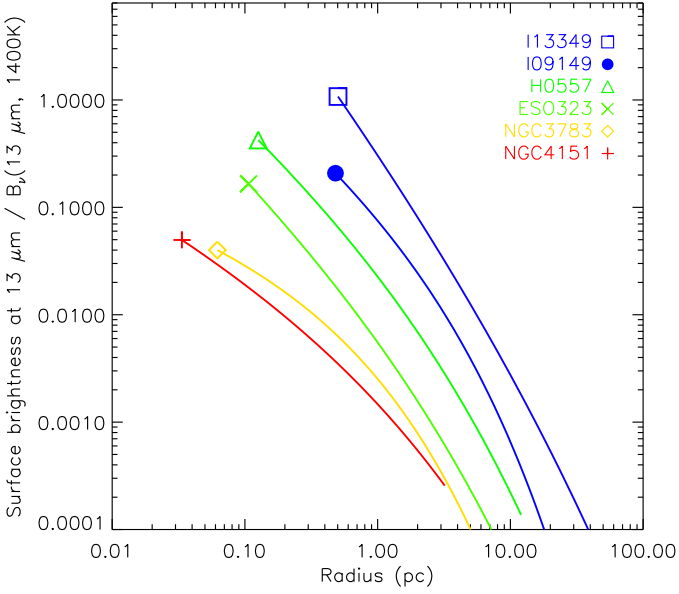}
\caption{The modeled surface brightness distribution of the power-law
  component at 13 $\mu$m as a
  function of radius in pc. The value is normalized with the constant
  value of the Planck function at 13 $\mu$m for $T$=1400K. The color
  indicates the UV luminosity of each object as in
  Fig.\ref{fig_all_vis_sfreq}, which illustrates that higher
  luminosity objects have steeper and brighter surface brightness.}
\label{fig_model_surfb}
\end{figure}

\begin{figure}
\centering \includegraphics[width=9cm]{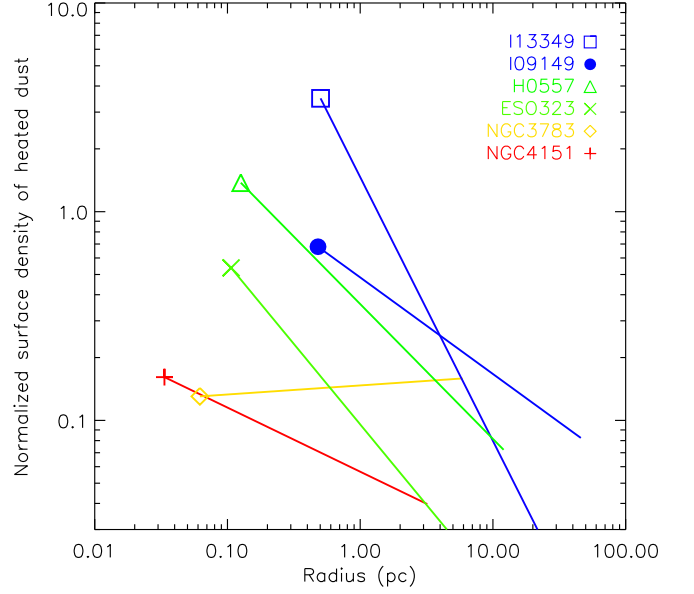}
\caption{The modeled surface density distribution of heated dust (at
  maximum temperature at each radius) in the
  power-law component, illustrating that higher
  luminosity objects have steeper and denser distributions.}
\label{fig_model_density}
\end{figure}

\subsubsection{Mapping the radial structure with a bright near-IR rim}

However, we would be able to roughly understand both the total flux
and size information if, in addition to the power-law structure, there
is a central {\Edit brightness} concentration on the scale of
$\sim$$\Rin$ emitting nearly at the sublimation temperature of
$\sim$1400K.  This concentration would correspond to a dust
sublimation region, and could be a bright rim of the innermost dust
distribution, such as the puffed-up rim that is often discussed for
the inner region of young stellar objects (see \citealt{Dullemond10}
for a review).  The exact structure of this central concentration is
not yet well constrained with our data.
Fig.\ref{fig_normnfn_rhalf_3panel}$c$ shows a grid of models with the
power-law plus the central concentration represented simply by a 1400K
ring at $r$=$\Rin$, which seems to closely reproduce the overall
SED-size relation with a range of density indices.  In this case, the
two objects, IRAS13349 and H0557-385, {\LE simply} have a slightly higher
flux level in the power-law component {\LE with respect to} the near-IR
ring.

In Figures \ref{fig_nfn_uv_sfreq_low} and \ref{fig_nfn_uv_sfreq_high},
we also show the best-fit results with this model for each object and
Table~\ref{tab_powerlaw} summarizes the parameters. The normalization
of the ring luminosity is parameterized in the same way as in the case
of Gaussian given in Eq.\ref{eq_gaussflux}, where $\Rhalf$ is replaced
by the radius of the ring.  Figure~\ref{fig_model_surfb} shows the
resulting surface brightness profiles of the models for each
object. The power-law component in higher luminosity objects becomes
steeper, as expected. Figure~\ref{fig_model_density} shows the
corresponding normalized surface density distribution of the heated
dust grains (i.e. with a maximum temperature at each radius).  It
becomes steeper, and also slightly denser, for higher luminosity
objects. The density indices are from $\sim$0 for lower luminosity to
$\sim$$-$1 for higher luminosity objects in our sample.

We note that the central temperature $\Tin$ of the mid-IR power-law
distribution has to be as low as $\sim$700 K, to reproduce the
relatively red mid-IR spectral shape.  (We fixed $\Tin$ to be 700K in
the fits implemented above, and we would need visibility data that
fill the gap between near-IR and mid-IR wavelengths to constrain
$\Tin$ more tightly.)  This would mean that there is a compact region
of such a low temperature quite close to the radius $\Rin$, which
could be the shadowed region just behind the bright sublimating rim.

\subsubsection{NGC4151}

In the case of NGC4151, this low-temperature compact region is
probably what we see in the mid-IR visibility curves, which become
relatively flat at spatial wavelengths smaller than $\sim$100~$\Rin$
as shown in the mid-panel of Fig.\ref{fig_nfn_uv_sfreq_low}$a$.  {\LE
  In} our interpretation, this is caused by both the high spatial
frequency part of the low-$\Tin$ power-law component and the hot
central brightness concentration. Here the effect of the latter on
visibility is indicated in this panel (also in the other mid-panels in
Fig.\ref{fig_nfn_uv_sfreq_low} and \ref{fig_nfn_uv_sfreq_high} for
other objects), where the dotted curves show the visibilities without
the bright near-IR rim.

These can also be understood in terms of the correlated flux observed
at long baselines (see the top-panel of
Fig.\ref{fig_nfn_uv_sfreq_low}$a$), which provides an approximate
spectrum of the unresolved part of the source. Its color temperature
is $\sim$400~K (or $\sim$500~K if we exclude slightly less reliable
data at wavelengths shortward of $\sim$10 $\mu$m), which is much
higher than that of the total flux ($\sim$280~K) and much lower than
the sublimation temperature.

The deviation of our model fit from the data might indicate that the
low-$\Tin$ core has even more flux while the outer radiating source is
even more extended.  This could {\LE suggest} a model of three
components (including the near-IR component), {\LE which was implied}
by \cite{Burtscher09}, who modeled their two-baseline mid-IR data with
a Gaussian and an unresolved source, and suggested another near-IR
component.  We note, however, that, puzzlingly, the correlated flux in
their two-baseline data {\LE did} not {\LE show the} relatively blue
color temperature, {\LE thus gave} almost no indication of the size
becoming smaller toward shorter wavelengths within the mid-IR.

\subsubsection{Warm mid-IR and hot near-IR parts}

The composite brightness structure discussed above, consisting of 
emission components of a warm mid-IR dominating part and a hot near-IR
dominating part, also seems to be discernible in the SED, particularly
in those of relatively red objects such as NGC3783 and NGC4151
(Fig.\ref{fig_nfn_uv_sfreq_low}).  The two-component explanation of
the near to mid-IR SEDs has been advocated by different authors from
SED analyses (e.g. \citealt{Mor09}). Historically, the bump around
3~$\mu$m in the SEDs of QSOs, which could correspond to the hot
near-IR part, has been known for decades (e.g. see sec.V in
\citealt{Neugebauer79}).

A direct size constraint {\LE on} the hot central brightness
concentration can be derived from near-IR interferometry, and its
structure might be {\LE a little distinct} from the warm power-law
distribution. For example, the near-IR visibility of NGC4151 suggests
that it has a quite compact structure with $\Rhalf$ quite close to
$\Rin$, while most of its mid-IR emission is quite extended.  {\LE In that
case}, another parameter, in addition to the heating luminosity, may
be responsible for forming the structure of the inner
region. We plan to explore these regions with further measurements
in both the near-IR and mid-IR.

\section{Conclusions}

We have presented mid-IR long-baseline interferometry of six type 1
AGNs.  Within the small sample, which nevertheless spans over
$\sim$2.5 orders of magnitudes in the UV luminosity $L$ of the central
engine, we have shown that the radial structure of the dust
distribution changes systematically with luminosity.  We have argued
that the mid-IR radial brightness distribution can approximately be
described by a power-law with the inner boundary set by the dust
sublimation radius $\Rin$. In this case, the power-law steepness can
be quantified with a half-light radius $\Rhalf$.  The mid-IR
half-light radius $\Rhalf$ in units of $\Rin$ becomes smaller
(i.e. the power-law distribution becomes steeper) as the luminosity
increases, with $\Rhalf$ at 13$\mu$m ranging from a few tens of $\Rin$
down to a few $\Rin$ over the range of $L$ from $\sim10^{43.5}$ to
$\sim 10^{46}$ erg/sec. This means that, in contrast to the results of
previous studies, the physical mid-IR radii in pc is {\it not}
proportional to $L^{1/2}$, but increases with $L$ much more slowly.
Our current estimate is that the radii is $\propto L^{+0.21\pm0.05}$ at 8.5
  $\mu$m, and nearly constant at 13~$\mu$m.

This mid-IR size normalized by $\Rin$ is correlated with the mid-IR
spectral shape, where the normalized size is smaller for bluer
spectral shapes. The mid- to near-IR SED as well as the measured size
in mid- and near-IR wavelengths can roughly be explained in a picture
where there is a central  brightness concentration at the scale
of $\sim$$\Rin$ emitting nearly at the dust sublimation temperature of
$\sim$1400K, over an underlying power-law-like structure extending
outwards.  The latter structure is inferred to have an innermost
temperature as low as $\sim$700K, and a radial surface
density distribution  of the heated dust that becomes steeper from
lower to higher luminosity objects, ranging from $\sim$$r^{0}$ to
$\sim$$r^{-1}$. The radial temperature {\LE run} is consistent with
$T\propto r^{\beta}$ of $\beta \sim-0.35\pm0.05$. The exact structure
of the innermost concentration will {\LE further} be constrained with the
planned near-IR interferometry.

\begin{acknowledgements}

  This research is primarily based on observations made with the
  European Southern Observatory telescopes obtained from the
  ESO/ST-ECF Science Archive Facility under the program 083.B-0452.
  Additional data sets were collected under the programs 082.B-0330
  and 083.B-0288.  The authors are grateful to Roy van Boekel for
  providing the spectral database of MIDI calibrators.  A part of the
  data presented here was obtained at the W.M. Keck Observatory, which
  is operated as a scientific partnership among the California
  Institute of Technology (Caltech), the University of California and
  the National Aeronautics and Space Administration (NASA). The
  Observatory was made possible by the generous financial support of
  the W.M. Keck Foundation.  The Keck Interferometer is funded by NASA
  as part of its Exoplanet Exploration program.  We are also grateful
  for the data obtained under the service observing program of the
  United Kingdom Infrared Telescope, which is operated by the Joint
  Astronomy Centre on behalf of the Science and Technology Facilities
  Council of the U.K.  This work is also based in part on observations
  made with the Spitzer Space Telescope, obtained from the NASA/ IPAC
  Infrared Science Archive, both of which are operated by the Jet
  Propulsion Laboratory, Caltech under a contract with NASA.  The
  research is also based in part on observations with AKARI, a JAXA
  project with the participation of ESA.  S.H. acknowledges support by
  Deutsche Forschungsgemeinschaft (DFG) in the framework of a research
  fellowship (``Auslandsstipendium'').  This work has made use of
  services produced by the NASA Exoplanet Science Institute at
  Caltech, and the NASA/IPAC Extragalactic Database (NED) which is
  operated by the Jet Propulsion Laboratory, Caltech, under contract
  with NASA. This research has also made use of the SIMBAD database,
  operated at CDS, Strasbourg, France.


\end{acknowledgements}



\begin{appendix}

\section{Reduction and calibration of MIDI data for faint targets}
\label{sec_software}

We used {\LE a part of} the software EWS (\citealt{Jaffe04SPIE};
version 1.7.1) to reduce the MIDI data described in this paper.  The
mid-IR correlated flux of all the targets here is $\lesssim$~0.5 Jy,
which is much fainter than normally handled with the
software. Therefore, we implemented several additional procedures to
reduce and calibrate the data as described below, using our own IDL
codes. In short, we simply averaged or smoothed over a relatively
large number of frames, designated as $w$ here (typically
$w\sim$20-40) to determine both the group-delay and phase-offset
tracks, and applied the same averaging {\LE to the} calibrator frames
to compensate for the side effects of the averaging.

We also observed several calibration stars, listed in
Table~\ref{tab_faintcal}, which are fainter than $\sim$0.8 Jy at 12
$\mu$m, have known spectral type, and are unresolved with
VLTI/MIDI. The deduced correlated flux is directly compared with the
total flux, which is obtained by scaling the template spectra of
\citealt{Cohen99IR} (not from less accurate total flux frames obtained
with MIDI). This facilitates a reliable evaluation of the accuracy of
our reduction and calibration procedures.  The scaling factors are
derived from both IRAS and AKARI measurements.  On the basis of the
close match between the two, we estimate that the total flux accuracy
is $\sim$4\%.

\subsection{Large gsmooth parameter}\label{sec_gsmooth}

When the group delay is determined from the delay function (the
Fourier transform of each fringe spectrum from the frequency domain to
the delay domain), several frames are averaged or frames are smoothed
{\LE over time direction} as specified by gsmooth parameter in EWS to
suppress instrumental delay peaks.  The default is four, but we used a
larger number (typically 10-20) to increase S/N in delay peak
determinations, as normally done for faint targets.

\subsection{Significant averaging over time for
  phase-offset determination}

Fig.\ref{figfcbright}(a) compares the calibrated correlated flux of
the unresolved star HD107485 with the known total flux spectrum
(i.e. the scaled spectrum described above), and
Fig.\ref{figfcbright}(b) shows the deduced visibility (using the
known total flux).  The raw count spectrum of the correlated flux is
shown in Fig.\ref{figfcbright}(c).  The star has a peak count rate
of $\sim$800 cts/sec in the PRISM-mode spectrum in this particular
observation. In these figures, the dotted line shows the spectrum from
a default reduction (though with the same large gsmooth parameter),
which determines phase offsets with a small averaging width.
In this case, the spectrum suffers a red bias, i.e. extra counts at
the longer wavelength side.

This {\Edit seems to be at least partly} due to the determination of
phase offsets at too low S/N.  This can be visualized on the complex
plane of the broad-band correlated flux counts integrated over the
N-band of $\sim$7 to 13 $\mu$m. The phase offsets are determined from
this integrated broad-band flux. When a large enough number of frames
are averaged, the distribution of the broad-band correlated flux
measurements on the complex plane shows a (partial) circle as shown in
Fig.\ref{figfcbright}(d), where its radius represents the
amplitude of the correlated flux with some phase offset slowly
changing over time.

A noisy phase-offset determination (and subsequent rotation using the
erroneous phase) would lead to a positive bias in the correlated flux
(e.g. it would detect non-zero correlated flux even if the real
correlated flux is zero).  The red bias is probably caused, since the
MIDI count spectra usually have a peak at shorter wavelengths, by the
phase-offset determination having a larger weight at shorter
wavelengths (or more accurately count peak), {\LE thus leading to} the
longer wavelength side having a stronger positive bias.  We note also
that, in general, correlated flux spectra would also tend to be redder
if larger coherence loss at shorter wavelengths is not correctly
calibrated.

Therefore, we chose a large $w$ for the phase offset track
determination, and a boxcar smoothing was applied to the group delay
track for the same number of frames. By applying the same averaging to
the delay and phase tracks of calibrator frames, we estimated the low
effective system visibility arising from this averaging process, and
calibrated it out from the target frames.  As we show below, we chose
$w$ to maximize the S/N of the phase-offset determination and the
result is insensitive to the exact value of the chosen $w$.

\subsection{Optimum averaging width and evaluation of atmospheric
  conditions}
\label{sec_maxsnr}

Here we quantify the effect of this averaging process as a function 
of $w$, and describe how we can choose an optimal value of~$w$.

\begin{figure*}
  \centering \includegraphics[width=\textwidth]
  {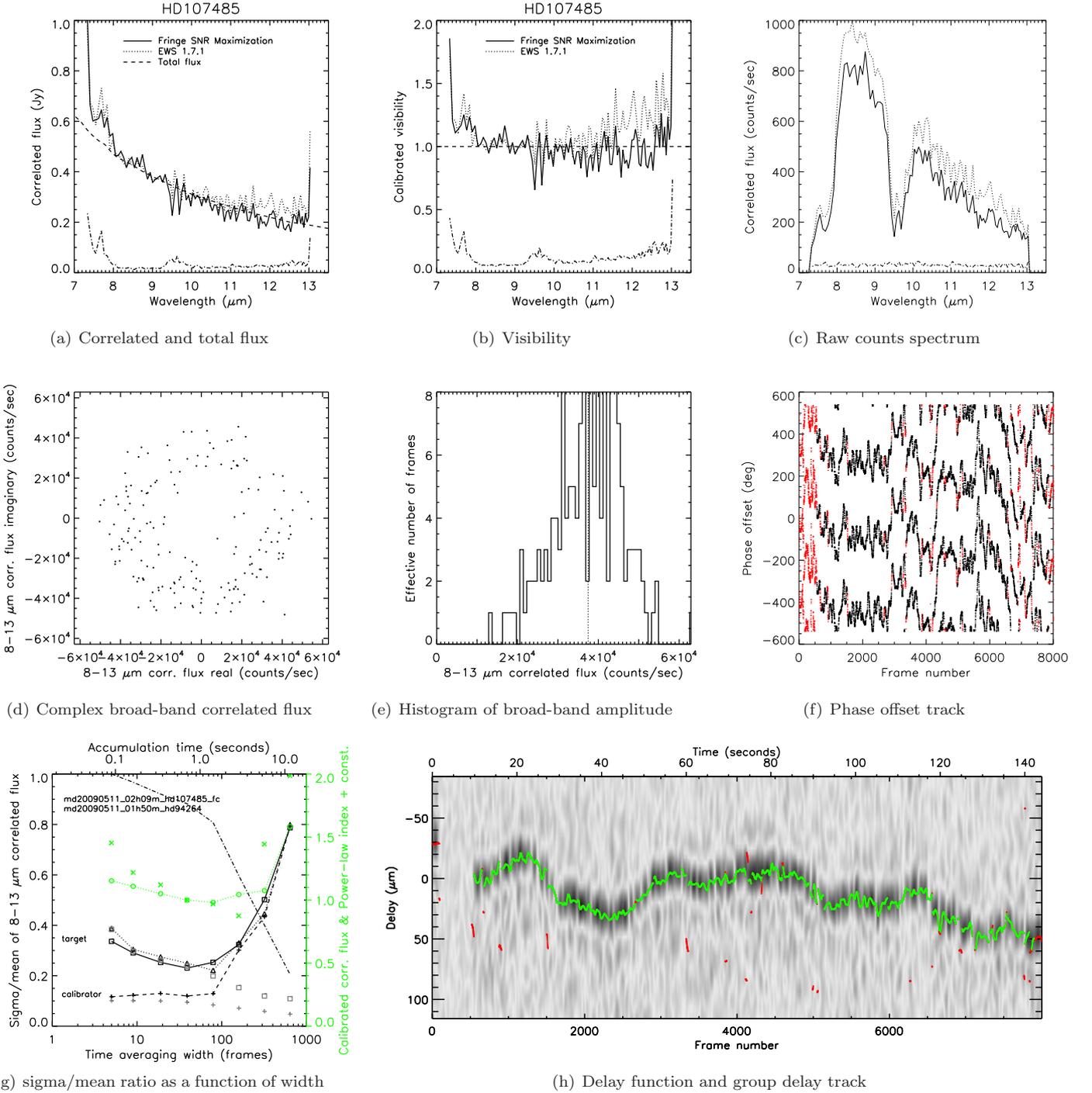}
  \caption{The results for the unresolved star HD107485 taken on
    2009-05-11. ($a$) Our results for the calibrated correlated flux
    and its error are shown in solid and dot-dash line, respectively,
    while the correlated flux from a default reduction with EWS 1.7.1
    (see text) is shown in dotted line. The dashed line is the total
    flux for this star which is obtained by scaling a template
    spectrum based on the known spectral type and IRAS flux. ($b$) The
    same as $a$ but deduced visibilities are shown instead of
    correlated flux. ($c$) The same as $a$ but raw counts spectra are
    shown. ($d$) Broad-band correlated flux measurements integrated
    over 7-13 $\mu$m (smoothed over a particular time width $w$) are
    shown on the complex plane. ($e$) Histogram of the amplitude of
    the broad-band correlated flux from the panel $d$ is shown. ($f$)
    The phase of the correlated flux as a function of frame number,
    i.e. time. The same track is repeatedly shown with 360$\degr$
    offsets to clearly indicate the track as a function of time. The
    red points show excluded frames. ($g$) The ratio of the standard
    deviation $\sigma$ to mean $m$ of the correlated flux amplitude
    measurements is shown as a function of smoothing width $w$ in
    frames (lower axis) and in seconds (upper axis). The solid line
    with squares is for the target (HD107485) while the dashed line
    with plus signs is for the calibrator (see
    Table~\ref{tab_faintcal}). The gray squares and gray plus signs
    (without any connecting lines) indicate the photon noise (see
    Sect.\ref{sec_maxsnr}) estimated for the target and calibrator,
    respectively. The dotted line with triangles shows the expected
    fractional dispersion $\sigma$/$m$ curve for the target, which is
    estimated from the $\sigma$/$m$ curve for the calibrator (i.e. the
    dahed line with plus signs) and the photon noise estimations
    (i.e. gray squares and plus signs with no connecting lines). The
    dash-dotted line indicates the relative correlated flux counts of
    the calibrator as a function of $w$, showing the relative
    effective system visibility at each $w$. The green circles show
    the relative calibrated broad-band correlated flux as a function
    of $w$. The green crosses show the relative power-law index of the
    calibrated correlated flux spectrum also as a function of
    $w$. ($h$) The delay function on the plane of delay versus frame
    number is drawn in gray scale image, showing the group delay track
    before the iteration described in Sect.\ref{sec_delay_refine}. The
    green and red points indicate non-rejected and rejected fringe
    peaks, respectively, after the iteration.}
  \label{figfcbright}
\end{figure*}

A set of broad-band correlated flux measurements for a given $w$ as
shown in the complex plane in Fig.\ref{figfcbright}(d) gives a
distribution of the correlated flux amplitude measurements. The
corresponding histogram is shown in Fig.\ref{figfcbright}(e). We
can evaluate the distribution as a function of $w$ by calculating a
fractional dispersion $\sigma/m$ (standard deviation $\sigma$ divided
by mean $m$) for each $w$. This statistical property would be valid as
long as the effective number of correlated flux measurements is not
low. Here, the effective number of correlated flux measurements is the
number of non-rejected frames (see section \ref{sec_delay_refine}
below) divided by $w$.  The reciprocal $m/\sigma$ corresponds to the
S/N of the correlated flux measurements per smoothed frame.

Fig.\ref{figfcbright}(g) shows the change of $\sigma/m$ as a
function of $w$ (or corresponding averaging time interval,
'accumulation time'; see upper axis) for the target and its visibility
calibrator.  The curve is cut when the effective number of
measurements becomes lower than 10. In the small $w$ regime for the
faint target, the dispersion naturally decreases with larger $w$ as
photon noise decreases.  The photon noise here comprises a background
Poisson noise and fluctuation due to background subtraction
residuals. This can be estimated from the noise spectra (a few hundred
frames with large OPD offsets, taken at the start of each fringe track
with MIDI) and the fractional fluctuation of the delay function
peaks. This is shown in gray squares in Fig.\ref{figfcbright}(g),
and closely matches the $\sigma/m$ curve at small $w$ range.  The
fractional dispersion $\sigma/m$ starts to increase at some width
where the effect of a large time-averaging width to smear and
effectively reduce the correlated flux becomes more dominant over the
photon noise decrease.

For the bright calibrator, $\sigma/m$ stays relatively constant at
small $w$ where flux fluctuation slowly decreases (indicated by gray
plus signs), while the smearing effect of averaging slowly
increases. The dash-dotted curve is the relative correlated flux
counts of the bright calibrator (normalized at the smallest $w$),
showing a relative effective system visibility at each $w$.  For
larger $w$, $\sigma/m$ increases quickly (system visibility declines
quickly) when the blurring effect of the large-$w$ averaging starts to
dominate. This gives an indication of the coherence time in the mid-IR
at the time of the observation ($\sim$1.5 seconds in the case of
Fig.\ref{figfcbright}(g)).  In the large $w$ regime, the
fractional dispersion $\sigma/m$ is thus determined predominantly by
the atmosphere at the time of the observation, rather than the
observed source.

The coherence time and thus the blurring effect is approximately the
same between target and calibrator frames when the two observations
take place close in time and on the sky.  In this case, we can
actually calculate the expected $\sigma/m$ for the target frames as a
function of $w$ from the calibrator frames.  We subtract the photon
noise (gray plus signs) in quadrature from the $\sigma/m$ of the
calibrator (solid plus signs with dashed lines) to obtain the
dispersion caused by averaging, and add it in quadrature to the photon
noise in target frames. This expected $\sigma/m$ curve for the target
is shown as triangle symbols with a dotted line in
Fig.\ref{figfcbright}(g), and closely matches the observed
$\sigma/m$, meaning that atmospheric conditions were very similar.  In
this case, we simply apply the same averaging width for the calibrator
frame to obtain an appropriate estimate of the effective system
visibility spectra, or transfer function over wavelengths, for the
target frames processed with this particular averaging width.

In Fig.\ref{figfcbright}(g), the green circles show the relative
calibrated broad-band correlated flux (normalized at $w$ giving
minimum $\sigma/m$, or maximum S/N) as a function of $w$, and the
green crosses show the relative power-law index of the calibrated
correlated flux spectrum. The calibrated correlated flux is
systematically higher, and often redder, at small $w$ because of the
positive bias from low S/N determination of phase offsets. We have
chosen $w$ to yield maximum S/N for the target. If $w$ is chosen to be
larger than this value, the result is quite insensitive to the exact
value of $w$, although the effective S/N of correlated flux
measurements is lower at larger $w$ owing to the smearing effect of
averaging.

\subsection{Adjustment for the atmospheric difference}
\label{sec_atmadj}

\begin{figure*}
  \centering \includegraphics[width=\textwidth]
  {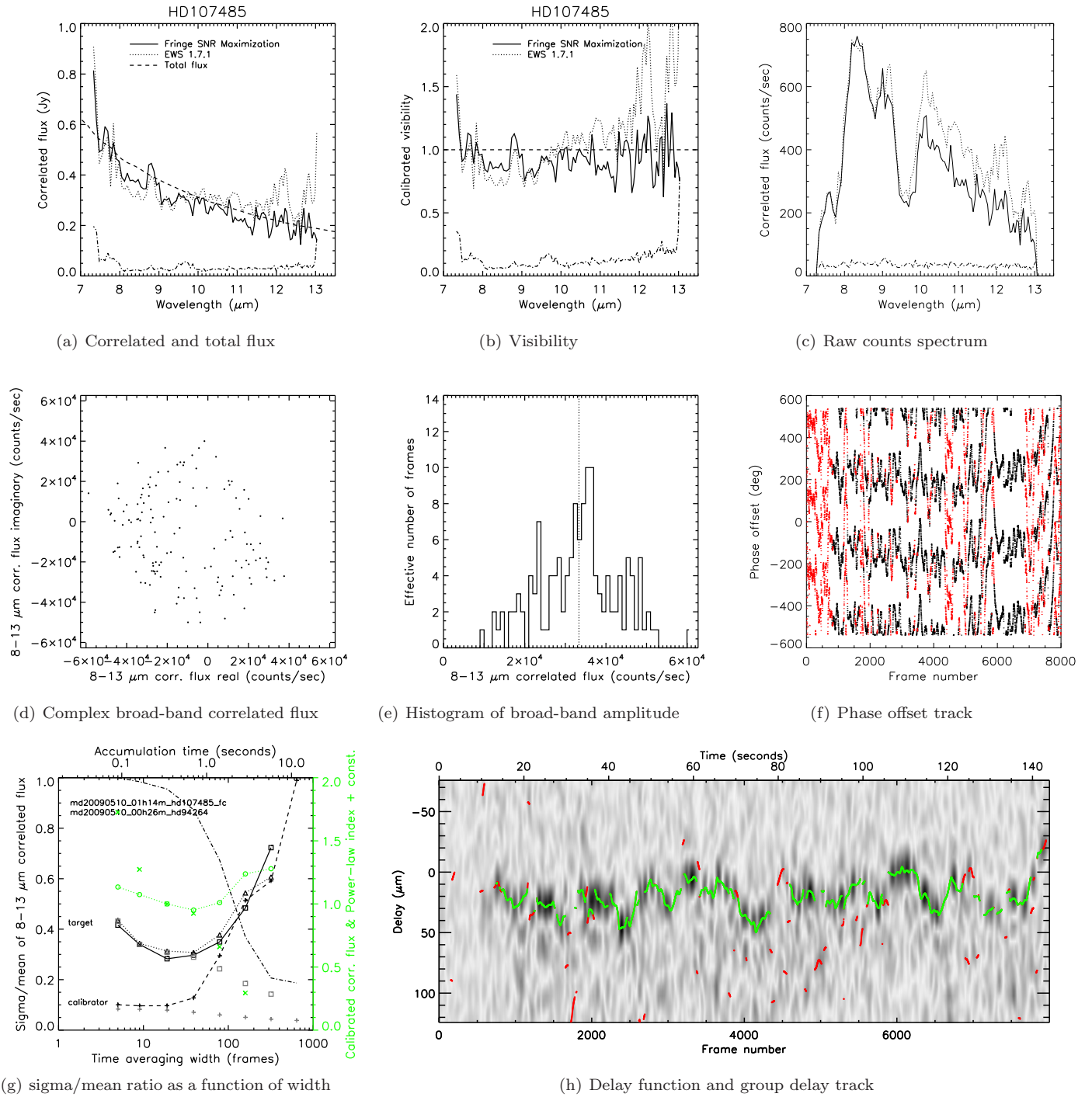}
  \caption{The same as Fig.\ref{figfcbright}, but for the data taken
    on 2009-05-10.}
  \label{figfcmedium}
\end{figure*}

Fig.\ref{figfcmedium} shows a similar case to that shown in
Fig.\ref{figfcbright}, where the count rate is slightly lower and
the coherence time is slightly shorter. The maximum S/N is thus
slightly lower.  However, the atmospheric conditions between target
and calibrator observations seem to be well matched.

In Figs.\ref{figfcfaint} and \ref{figfcfaint2}, we show the cases where
there seems to be a difference in atmospheric conditions between
target and calibrator fringe tracks.  In these cases, we try to
evaluate the difference by comparing the expected $\sigma/m$ curve for
the target (triangles with dotted line in
Fig.\ref{figfcfaint}(i) and \ref{figfcfaint2}(i)),
as derived from the calibrator frames, with the observed curve for the
target (squares with solid lines). We find that we can fit these two
curves if we simply scale the time length of the calibrator frames (a
factor of $\sim$0.7 in Fig.\ref{figfcfaint}(g), and $\sim$2.5 in
Fig.\ref{figfcfaint2}(g)).  The scaling factor probably corresponds
to the difference in the coherence time between target and calibrator
frames.  Therefore, we infer that we can obtain appropriate
calibrations of the target frames if we use this time scaling factor.
This correction also stabilizes the calibrated correlated flux over
different $w$.


Fig.\ref{figfccomp} shows the calibrated visibility of the 12 low and
intermediate S/N observations of unresolved stars as a function of the
maximum S/N per smoothed frame. We calculated the visibility in three
different wavelength bins, avoiding the range affected by atmospheric
ozone features.  We confirm that the calibration procedure above
correctly calibrates the data, although we see a slight tendency that,
toward shorter wavelengths, the correlated flux is underestimated as
the maximum S/N decreases. Based on the mean deviation and dispersion
of these calibrated visibilities, we assign a systematic error in the
correlated flux {\LE as} 5\% of the total flux, when the maximum S/N
is at least larger than $\sim$2.5.  Since the cases at lower S/N
remain untested, we did not use the target datasets under this level
in this paper.

\subsection{Refinement of group delay determination}
\label{sec_delay_refine}

As described in section~\ref{sec_gsmooth}, we use a large gsmooth
parameter for faint targets to smooth over a large number of delay
function frames and thus increase the S/N for delay determination and
also strongly suppress the instrumental delay peaks.  However, this also
suppresses the atmospheric delay peaks at the time intervals where the
delay is relatively quickly changing. As a result, the delay function
image becomes quite 'blobby', and strong delay peaks are left only
where the atmospheric delay did not change rapidly.

\begin{figure*}
  \centering \includegraphics[width=\textwidth]
  {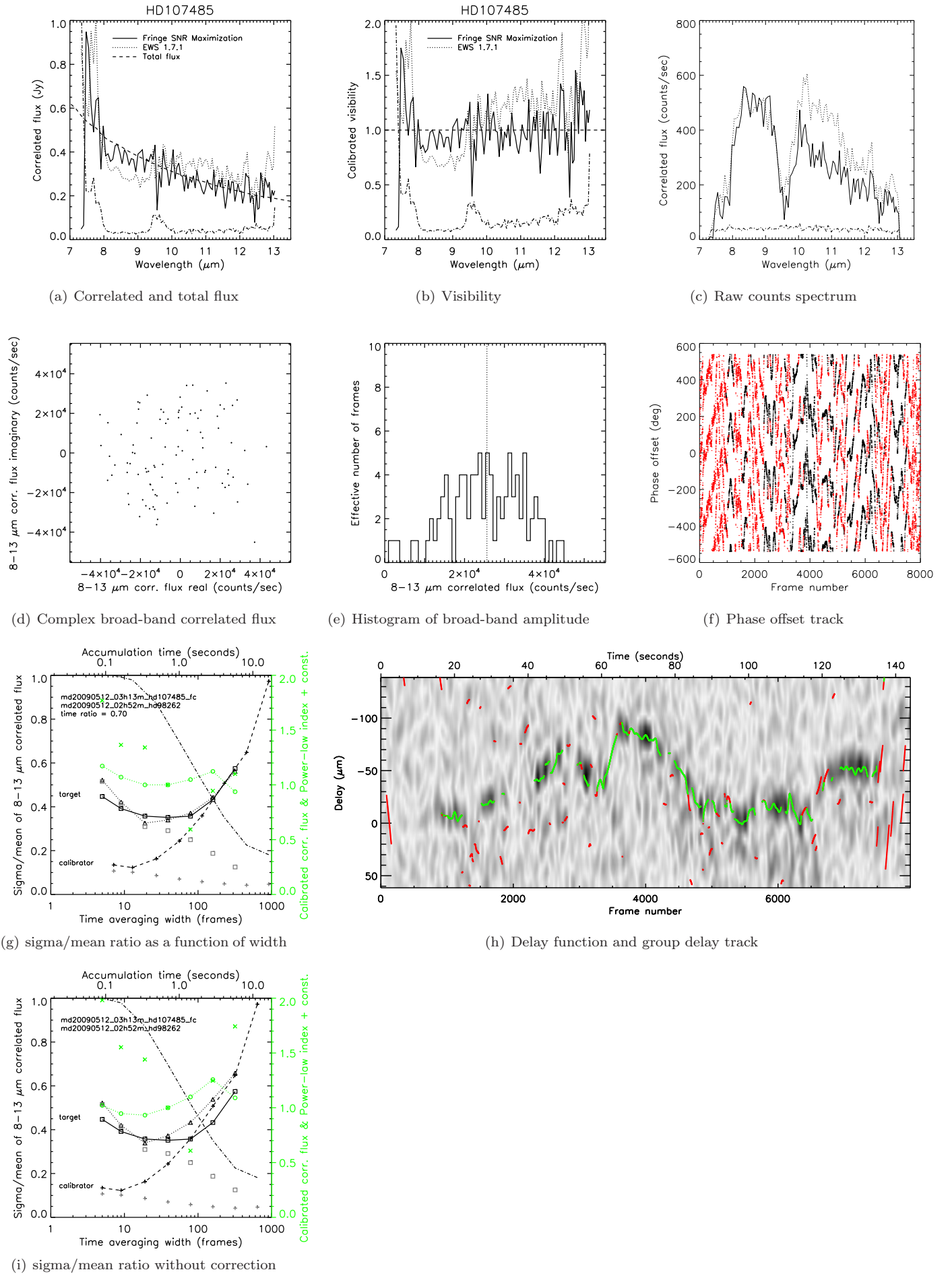}
  \caption{The same as Fig.\ref{figfcbright}, but for the data taken
    on 2009-05-12. The panel ($g$) shows the $\sigma$/$m$ curve after
    the correction for the time scale difference between the target
    and calibrator, while the panel ($i$) shows the curves before the
    correction.}
  \label{figfcfaint}
\end{figure*}

\begin{figure*}
  \centering \includegraphics[width=\textwidth]
  {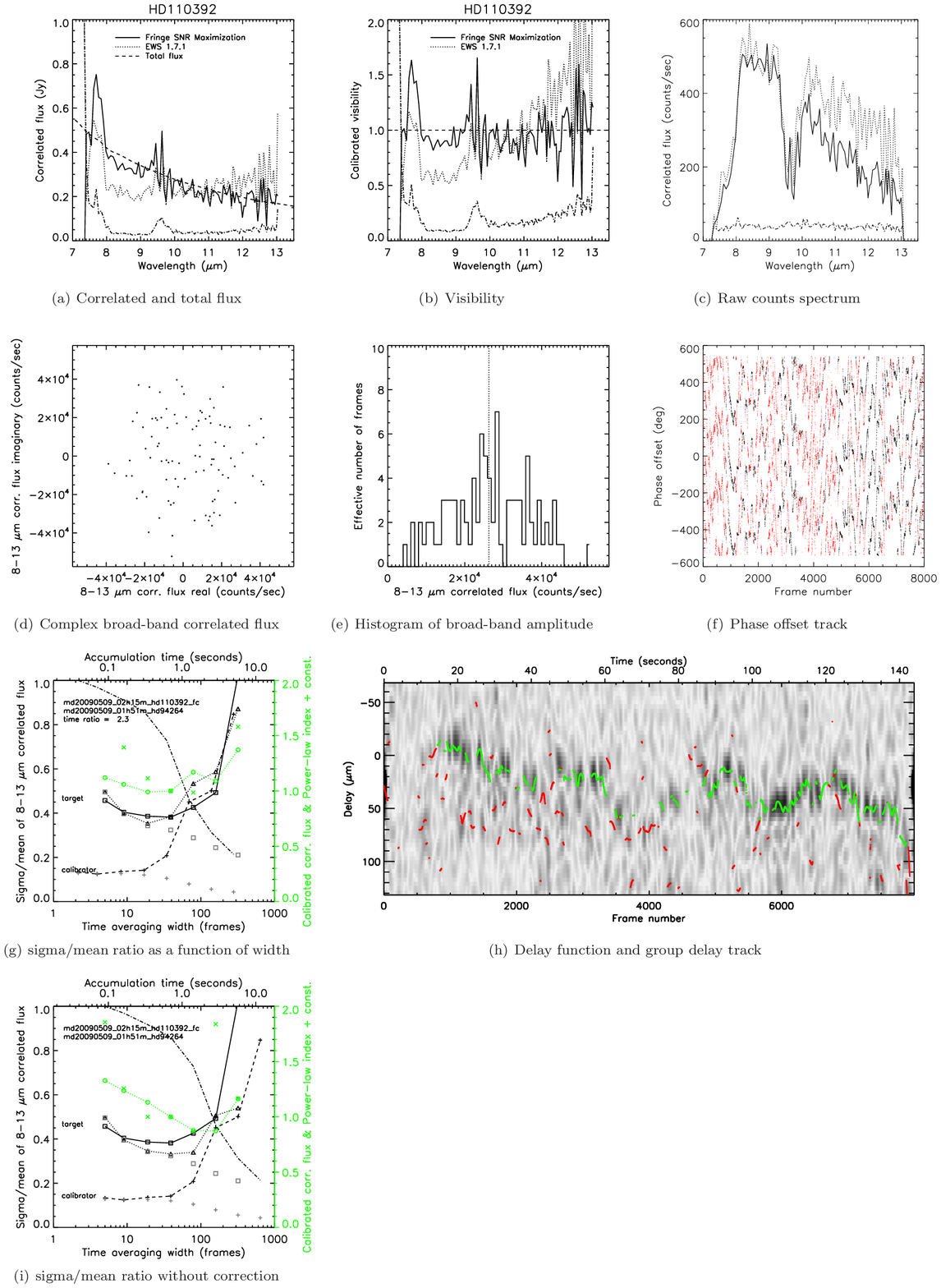}
  \caption{The same as Fig.\ref{figfcfaint} but
  for another unresolved star HD110392 taken on 2009-05-09.}
  \label{figfcfaint2}
\end{figure*}

\begin{figure}
\centering \includegraphics[width=9cm]{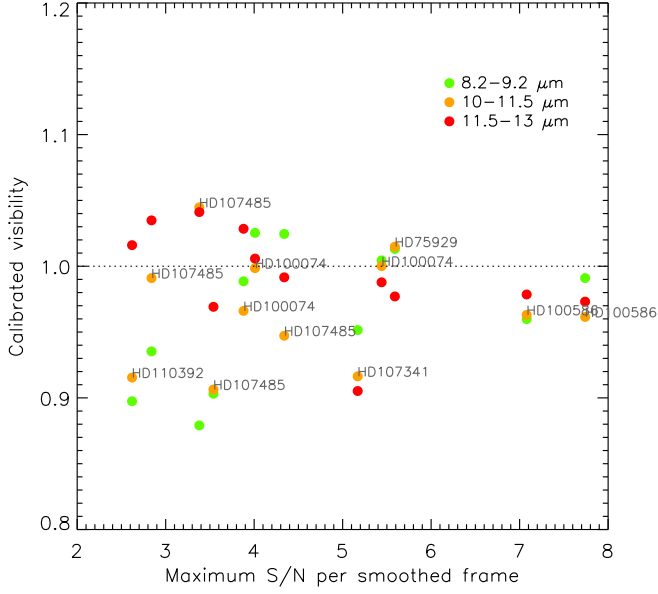}
\caption{Comparison of deduced visibilities for sub-Jy calibrators.}
\label{figfccomp}
\end{figure}

To partially compensate for this effect, we implemented a simple
iteration to determine the delay track. Namely, we first interpolated
(over time) between the strong delay peaks to derive an approximate
delay track.  Then we used this approximate atmospheric delay track
(plus instrumental delay) to rotate the original fringe spectra in the
complex plane, and re-derived the delay function smoothed with the
same gsmooth parameter. This makes the resulting delay peaks line up
straighter over time, and thus recovers a larger number of frames
with strong delay peaks. We cycled over this process a few times. For
calibrator frames, this resulted in almost all the frames having
strong peaks (i.e. even with the large gsmooth parameter).  For target
frames, significant fraction of frames is recovered, resulting in a
good representation of the delay track even in a relatively low S/N
case as shown in Fig.\ref{figfcfaint}.

In low S/N cases, we flagged frames that show delay peaks at
positions that deviate significantly from the overall track obtained
above. This was done by taking the histogram of the difference between
the determined group delay track with that smoothed over many frames
(typically $\sim$20-40 frames, set as 2 $\times$ gmoooth). The
distribution can roughly be descibed by a Gaussian, and frames with
large deviations ($>$ 10 $\sigma$) were excluded.  In this rejection
process, we did not select frames based on the strength of the delay
peaks in order to avoid possible biasing.


\subsection{A note on total flux frames}

For the targets observed with UTs that have total flux of less than a
few Jy, it is better to implement an additional background subtraction
using the sky regions very close to the target position. The residual
of the primary background subtraction using chopped frames can still
dominate the target flux for these relatively faint targets.  An
optional way to further reduce the systematic error from the
background subtraction is to iteratively reject frames with the sky
region counts having large deviations from the average.  This was
implemented in a small number of cases when the time fluctuation over
the photmetry frames was significantly reduced after the frame
rejection.  With this additional sky subtraction, we averaged many
sets of data, even from different nights, to obtain reliable total
flux spectra, as confirmed with our VISIR photometry and spectrum.


\begin{table}

\caption[]{Observation log for sub-Jy unresolved stars.}
\begin{tabular}{lcccccccccccccc}
\hline
name & flux     & \multicolumn{2}{c}{date \& time} & vis./flux & max   \\ 
     & (Jy)$^a$ & \multicolumn{2}{c}{(UT)}         & calibrator& S/N$^b$\\ 
\hline
HD100586 & 0.82 & 2009-03-13 & 04:44 & HD101666 & 7.1\\ 
         &      & 2009-03-15 & 04:00 & HD101666 & 7.7\\ 
HD107341 & 0.75 & 2009-05-09 & 03:34 & HD98262  & 5.2\\  
HD75929  & 0.55 & 2009-03-15 & 01:02 & HD100407 & 5.6\\ 
HD100074 & 0.45 & 2009-03-14 & 04:17 & HD100407 & 5.4\\ 
         &      & 2009-03-15 & 02:19 & HD100407 & 4.0\\ 
         &      & 2009-05-10 & 03:38 & HD112213 & 3.9\\ 
HD107485 & 0.34 & 2009-05-10 & 01:14 & HD94264  & 3.5\\  
         &      & 2009-05-11 & 02:09 & HD94264  & 4.3\\  
         &      & 2009-05-12 & 03:06 & HD98262  & 3.4\\  
         &      & 2009-05-12 & 03:13 & HD98262  & 2.8\\  
HD110392 & 0.28 & 2009-05-09 & 02:15 & HD94264  & 2.6\\  
\hline
\end{tabular}
\\
$^a$ 12 $\mu$m Flux from IRAS Faint Source Catalog.\\
$^b$ Maximum signal-to-noise ratio per smoothed frame for the
correlated\\ flux integraged over the whole N-band.

\label{tab_faintcal}
\end{table}

\section{Mid-IR data plotted in conventional formats}
\label{sec_conv_format}

In Fig \ref{fig_nfn_uv_sfreq_low} and \ref{fig_nfn_uv_sfreq_high} we
show our flux and visibility data with normalizations using the inner
radius $\Rin$. We show the same data in the mid-IR (after reddening
corrections; see Sect.\ref{sec_deredden}) using conventional
units in Fig \ref{fig_fnu_vis_low} and \ref{fig_fnu_vis_high}, to facilitate
direct comparisons with the data in the literature.

\renewcommand{\miniwidth}{0.3\textwidth}
\renewcommand{\figwidth}{6.5cm}
\renewcommand{\figintvl}{0.03\textwidth}

\renewcommand{\figwidth}{6.3cm}

\renewcommand{\froot}{fnu_vis_}
\renewcommand{\labelroot}{fig_fnu_vis_}

\renewcommand{\objnamea}{NGC4151}
\renewcommand{\objnameb}{NGC3783}
\renewcommand{\objnamec}{ESO323-G77}

\begin{figure*}
\input{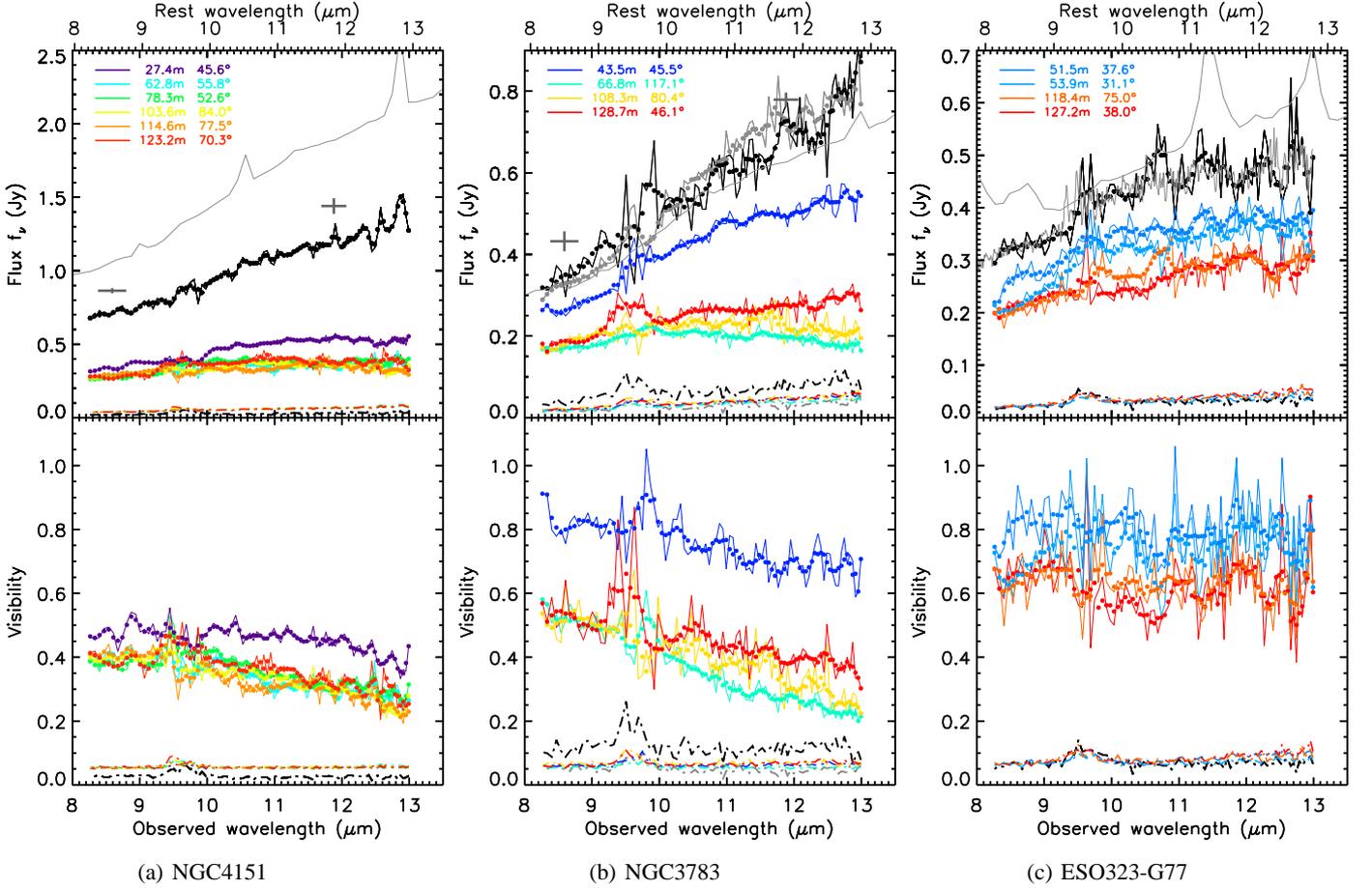}
\caption{ All the mid-IR data shown in Figs~\ref{fig_nfn_uv_sfreq_low}
  and \ref{fig_nfn_uv_sfreq_high} and summarized in
  Table~\ref{tab_summary} are plotted here in more conventional ways
  for (a) NGC4151, (b) NGC3783, and (c) ESO323-G77.  {\it Top}: the
  total flux and correlated flux spectra taken with MIDI are shown in
  units of Jy. The filled circles indicate the spectra with a 5-pixel
  boxcar smoothing (3 pixel for NGC4151), while the spectra before the
  smoothing are shown with thin lines. The total flux is shown in
  black, while the correlated flux is shown in different colors for
  different baselines as indicated by the legend at top-left where the
  projected baseline lengths and PAs are written. The error spectra
  estimated for the smoothed spectra are shown at the bottom with
  dash-dot lines using the same corresponding colors.  The two fluxes
  from the broad-band VISIR imaging are shown with error bars in thick
  gray lines. The thin gray spectra mostly at the highest flux levels
  are from the Spitzer IRS observations. The slightly darker gray
  spectrum shown for ESO323-G77 is from the VISIR spectroscopy. For
  NGC3783, the total flux taken in 2005 scaled to match the flux in
  2009 (see Sect.\ref{sec_midi}) is shown in gray color.  {\it
    Bottom}: the resulting visibility spectra are shown using the same
  color scheme, with the error spectra excluding the total flux error
  contribution. The fractional total flux errors are drawn separately
  in black dash-dot lines.  }

\label{\labelroot low}
\end{figure*}

\renewcommand{\objnamea}{H0557-385}
\renewcommand{\objnameb}{IRAS09149}
\renewcommand{\objnamec}{IRAS13349}

\begin{figure*}
\input{fig_nfn_uv_sfreq}
\caption{The same as Fig.\ref{fig_fnu_vis_low}, but the data shown in
  solid lines are with the total flux smoothed by 3 pixels while those
  in filled circles are with total and correlated fluxes smoothed by 7
  pixels. The figures are for (a) H0557-385, (b) IRAS09149-6206, and
  (c) IRAS13349+2438. }
\label{\labelroot high}
\end{figure*}

\section{Radial structure plot in linear scale}\label{sec_linear}

\begin{figure*}
\centering \includegraphics[width=9cm]{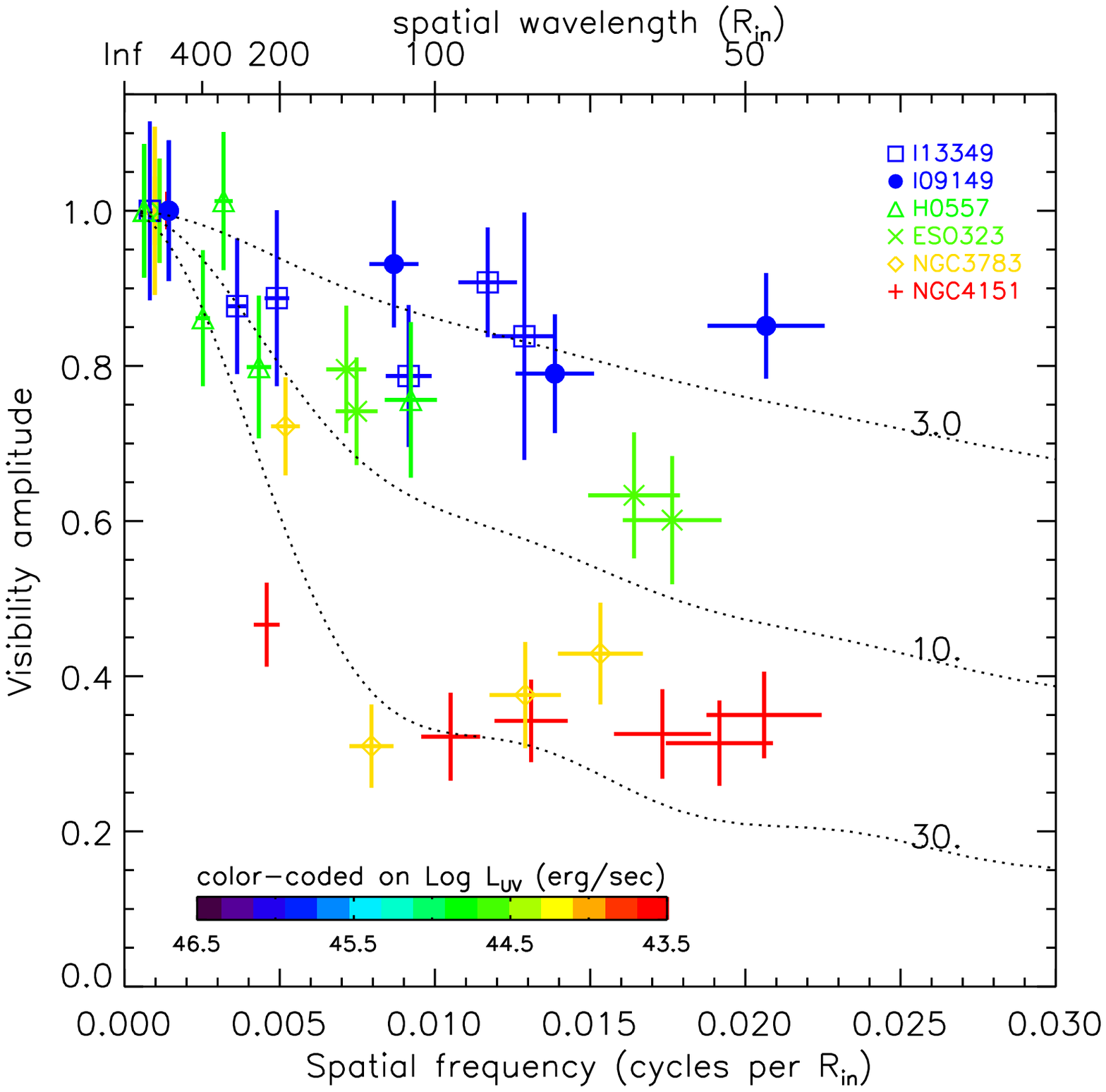}
\caption{The same figure as Fig.\ref{fig_all_vis_sfreq}, but 
with the spatial frequency on a linear scale.}
\label{fig_all_vis_sfreq_linear}
\end{figure*}

In Fig.\ref{fig_all_vis_sfreq} we showed observed mid-IR visibilities
of all the targets in a single plot to uniformly compare the radial
structure of the objects in normalized units with spatial frequency in
log scale. Here in Fig.\ref{fig_all_vis_sfreq_linear} we show exactly
the same figure but with spatial frequency in linear scale, again to
facilitate comparisons with the data in the literature.

\end{appendix}

\end{document}